\documentclass{aa}  

\usepackage{graphicx}
\usepackage{txfonts}

\newcommand\rout{R_{\mathrm{out}}}
\newcommand\rin{R_{\mathrm{in}}}
\newcommand\dsub{{\mathrm{d}}}
\newcommand\asub{{\mathrm{a}}}

\begin{document}

   \title{Testing eccentricity pumping mechanisms to model eccentric long period sdB binaries with MESA}

  
  \author{
    J.~Vos
    \inst{1}
    \and
    R.H.~\O{}stensen
    \inst{1}
    \and
    P.~Marchant
    \inst{2}
    \and
    H.~Van~Winckel
    \inst{1}
   }
          
  \institute{
    Instituut voor Sterrenkunde, KU Leuven, Celestijnenlaan 200D, B-3001 Leuven, Belgium\\
    \email{joris.vos@ster.kuleuven.be}
    \and
    Argelander-Institut für Astronomie, Universität Bonn Auf dem Hügel 71, D-53121 Bonn, Germany 
    }

   \date{Received \today; accepted ???}

 
  \abstract
   {Hot subdwarf-B stars in long-period binaries are found to be on eccentric orbits, even though current binary-evolution theory predicts those objects to be circularised before the onset of Roche-lobe overflow (RLOF).}
   {We aim to find binary-evolution mechanisms that can explain these eccentric long-period orbits, and reproduce the currently observed period-eccentricity diagram.}
   {Three different processes are considered; tidally-enhanced wind mass-loss, phase-dependent RLOF on eccentric orbits and the interaction between a circumbinary disk and the binary. The binary module of the stellar-evolution code MESA (Modules for Experiments in Stellar Astrophysics) is extended to include the eccentricity-pumping processes. The effects of different input parameters on the final period and eccentricity of a binary-evolution model are tested with MESA.}
   {The end products of models with only tidally-enhanced wind mass-loss can indeed be eccentric, but these models need to lose too much mass, and invariably end up with a helium white dwarf that is too light to ignite helium. Within the tested parameter space, no sdBs in eccentric systems are formed. Phase-dependent RLOF can reintroduce eccentricity during RLOF, and could help to populate the short-period part of the period-eccentricity diagram. When phase-dependent RLOF is combined with eccentricity pumping via a circumbinary disk, the higher eccentricities can be reached as well. A remaining problem is that these models favour a distribution of higher eccentricities at lower periods, while the observed systems show the opposite.}
   {The models presented here are potentially capable of explaining the period-eccentricity distribution of long-period sdB binaries, but further theoretical work on the physical mechanisms is necessary.}


   \maketitle
%

\section{Introduction}
Hot subdwarf-B (sdB) stars are core-helium-burning stars with a very thin hydrogen envelope (M$_{\rm{H}}$ $<$ 0.02 $M_{\odot}$), and a mass close to the core-helium-flash mass $\sim$ 0.47 $M_{\odot}$ \citep{Saffer94, Brassard01}. These hot subdwarfs are found in all galactic populations, and they are the main source for the UV-upturn in early-type galaxies \citep{Green86,Greggio90, Brown97}. Furthermore, their photospheric chemical composition is governed by diffusion processes causing strong He-depletion and other chemical peculiarities \citep{Heber98}. 
The formation of these extreme-horizontal-branch objects is still puzzling. To form an sdB star, its progenitor needs to lose its hydrogen envelope almost completely before reaching the tip of the red-giant branch (RGB), so that the core ignites while the remaining hydrogen envelope is not massive enough to sustain hydrogen-shell burning. A variety of possible formation channels have been proposed. 
Currently, there is a consensus that sdB stars are only formed in binaries. Several evolutionary channels have been proposed, where binary-interaction physics plays a major role. Close binary systems can be formed in a common-envelope (CE) ejection channel \citep{Paczynski76}, while stable Roche-lobe overflow (RLOF) can produce wide sdB binaries \citep{Han00, Han02}. An alternative formation channel forming a single sdB star is the double white-dwarf (WD) merger, where a pair of white dwarfs spiral in to form a single sdB star \citep{Webbink84}.The possibility of forming a single sdB after a merger with a sub-stellar object has also been proposed \citep{Soker2014}, but not modelled. This channel could produce sdBs with a narrow mass distribution unlike the WD mergers.

\citet{Han02, Han03} addressed these three binary-formation mechanisms, and performed binary-population-synthesis (BPS) studies for two kinds of CE-ejection channels, two possible stable-RLOF channels and the WD-merger channel. The C- ejection channels produce close binaries with periods of $P_{\rm{orb}}$ = 0.1 -- 10 d, and main-sequence (MS) or white-dwarf (WD) companions. The sdB binaries formed through stable RLOF have orbital periods ranging from 10 to 500 days, and MS companions. \citet{Chen13} revisited the RLOF models of \citet{Han03} with more sophisticated treatment of angular-momentum loss. When including atmospheric RLOF, these revised models can reach orbital periods as long as $\sim$1600\ d. Finally, The WD-merger channel can lead to single sdB stars with a higher mass, up to 0.65 $M_{\odot}$.  A detailed review of hot subdwarf stars is given by \citet{Heber09}.

Many observational studies have focused on short-period sdB binaries \citep{Koen98,Maxted00,Maxted01,Heber02,Morales03,Napiwotzki04,Copperwheat11}, and over 100 of these systems are currently known \citep[Appendix A]{Geier11}. These observed short-period sdB binaries correspond very well with the results of BPS studies. Currently, only nine long-period sdB binaries are known \citep{Green01,Oestensen2011,Oestensen2012,Deca12,Barlow12,Vos12,Vos13,Vos14}. Even though this is a small sample, their period-eccentricity distribution is not compliant to the prediction of evolution models. All current models predict circular orbits, while seven out of nine observed systems have a significantly eccentric orbit. 

This eccentricity problem is not entirely new, and a few possible solutions have been proposed, although none have been applied to the case of sdB binaries. There are three potential mechanisms that can create eccentric orbits described in the literature. \citet{Soker2000} based on theoretical work of \citet{Eggleton2006} proposed the mechanism of tidally-enhanced wind mass-loss to allow the eccentricity of the orbit to increase. In this framework, the wind mass-loss is increased due to the tidal influence of a companion in an eccentric orbit, and the effect is present before the system comes into contact. It reduces the tidal forces by keeping the sdB progenitor within its Roche lobe, while the phase-dependent wind mass-loss can increase the eccentricity. This mechanism is also known as Companion Reinforced Attrition Process (CRAP). \citet{Siess2014} used this mechanism to successfully explain the orbit of the highly eccentric He-WD binary IP Eri.

A second possible mechanism is that of phase-dependent mass loss during RLOF. Similar to the phase-dependent mass loss in the CRAP mechanism, a varying mass-loss rate in a binary that is larger during periastron than apastron, can increase the eccentricity of the orbit. The difference with the previous mechanism is that it is active during RLOF. This mechanism was used by \citet{Bonacic08}, who used a model with enhanced mass loss from the AGB star due to tidal interactions and a smooth transition between the wind mass loss and the RLOF mass loss regimes to explain the eccentric population of post-AGB binaries.

The third method of increasing the eccentricity of binaries is by interaction of the binary with a circumbinary (CB) disk. The motivation to include this process in the context of sdB stars, is the observational finding that stable circumbinary disks are commonly observed around evolved post-AGB binaries \citep[e.g.][and references therein]{deRuyter2006, Hillen2014}. The longevity of these disks are corroborated by the strong processing of the dust grains as attested by the infrared spectral dust-emission features \citep[e.g.][]{Gielen2008, Gielen2011} and the millimetre continuum fluxes that indicate the presence of large grains \citep{deRuyter2005}, while interferometric techniques are needed to resolve them \citep[see e.g.][and references therein]{Hillen2014}. 
The Keplerian rotation is, up to now, only spatially resolved in two objects \citep{Bujarrabal2005, Bujarrabal2015}. Recent surveys of the Large and Small Magellanic Clouds \citep{vanAarle2011, Kamath2014} show that a significant fraction of post-AGB stars show the distinctive near-IR excess indicative of a stable disk. One of the results of \citet{Kamath2014, Kamath2015} is that a significant population of post-RGB stars were identified with circumstellar dust likely in a disk as well. The disk evolution will determine the infrared life time of the systems and hence the detectability. While there is ample observational evidence that disks are common in evolved binaries, we assume here that disks were also present during the RGB evolution of the sdB progenitor. 
\citet{Dermine2013} explored the eccentricity-pumping mechanisms of CB disks in post-AGB binaries, based on theoretical work and smooth-particle-hydrodynamic (SPH) simulations of \citet{Artymowicz1994} and \citet{Lubow1996}. An issue with the results of \citet{Dermine2013} is that there was a mismatch between the disk mass distributions used to derive different parts of the interaction mechanisms. 

In this article we will explore these three methods to find formation channels that can explain the eccentricity of sdB binaries, using the stellar/binary evolution code MESA. In the case of CB disks that are formed during RLOF, the eccentricity pumping effect of the disk is combined with the effect of phase-dependent RLOF. We do not aim to perform a binary-population-synthesis study, but to explore in a limited number of initial conditions, the efficiency of the different processes. Our aim is to describe the effect of the model parameters on the final period and eccentricity of the binary, and to discover which areas in the period-eccentricity diagram can be covered by the three methods. 

An overview of the currently known systems is given in Sect. \ref{sec_observations}. The MESA evolution code is explained in Sect. \ref{sec_mesa}. The modelling methodology is explained in Sect.\,\ref{sec_methodology}. While the different models are presented in Sects. \ref{sec_crap}, \ref{sec_rlof} and \ref{sec_disk} for respectively the CRAP models, phase-dependent RLOF models and models containing a CB disk. The obtained period-eccentricity distribution is discussed in Sect. \ref{sec_period_ecc_distribution}, and a summary and conclusion is given in Sect. \ref{sec_conclusion}. A detailed overview of the binary physics used in MESA is given in appendix \ref{app_mesa}.

\section{Observed period-eccentricity diagram}\label{sec_observations}
\begin{figure}[!t]
\centering
\includegraphics{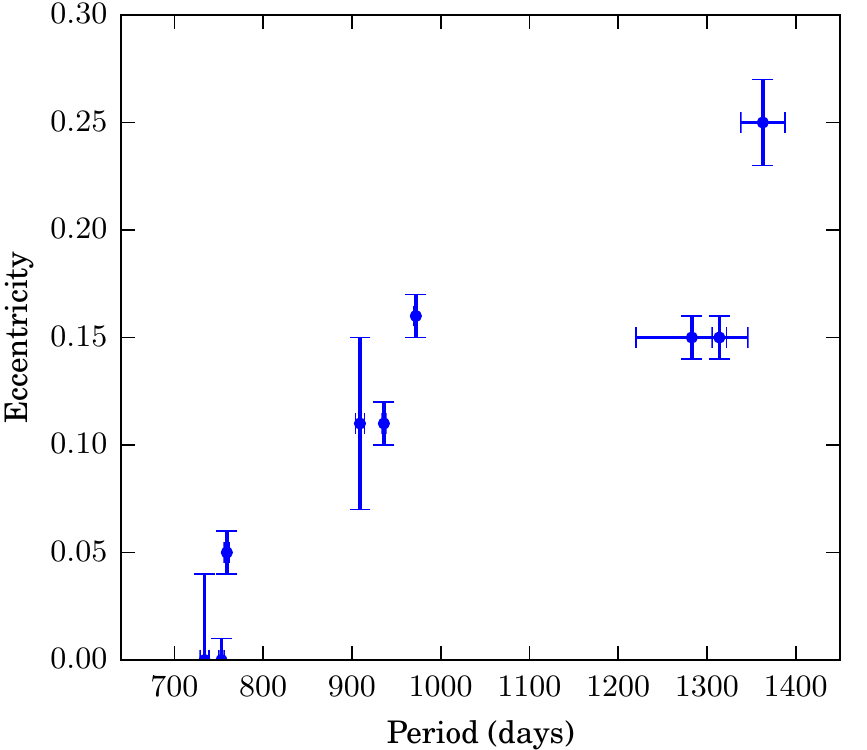}
\caption{The observed period-eccentricity diagram of all known long-period sdB binaries. There is a clear trend visible of higher eccentricities at higher orbital periods. See also Table \ref{tb_observed_period_ecc}.}\label{fig_observed_period_ecc}
\end{figure}

There are currently two wide sdB binaries with circular orbits and seven significantly eccentric systems covered in the literature. The first observed long-period sdB binary was PG\,1018$-$047 with a period of 760 $\pm$ 6 days and an unclear eccentricity \citep{Deca12}. Based on new observations \citet{Deca2015} revised this to a period of 759 $\pm$ 3 days and an eccentricity of 0.05 $\pm$ 0.01. \citet{Barlow12} published the orbits of two sdB systems, PG\,1701$+$359 and PG\,1449$+$653, the first circular and the second with an eccentricity of 0.11$\pm$0.04. There is one extra circular system known, PG\,1104$+$243 \citep{Vos12}. Furthermore, five eccentric systems were observed with the Mercator telescope \citep{Vos13, Vos14} using the HERMES spectrograph \citep{Raskin2011}. 
All known orbits are given in Table \ref{tb_observed_period_ecc}, while the resulting period-eccentricity diagram is shown in Fig. \ref{fig_observed_period_ecc}. Even though it is based on published data, this distribution has not been shown this clearly before.

\begin{table}
\caption{The observed periods and eccentricities of all known long-period sdB binaries. See also Fig. \ref{fig_observed_period_ecc}}\label{tb_observed_period_ecc}
\centering
\begin{tabular}{lr@{ $\pm$ }lr@{ $\pm$ }l}
\hline\hline
\noalign{\smallskip}
Object name	&	\multicolumn{2}{c}{Period (days)}	&	\multicolumn{2}{c}{Eccentricity}\\\hline
\noalign{\smallskip}
PG\,1701$+$359\tablefootmark{a}         &  734    &   5   &  0.00   &   0.04   \\
PG\,1104$+$243\tablefootmark{b}         &  753    &   3   &  0.00   &   0.01   \\
PG\,1018$+$243\tablefootmark{c}         &  759    &   3   &  0.05   &   0.01   \\
PG\,1449$+$653\tablefootmark{a}         &  909    &   5   &  0.11   &   0.04   \\
Feige\,87   \tablefootmark{d}           &  936    &   2   &  0.11   &   0.01   \\
BD$+$34$^{\circ}$1543\tablefootmark{d}  &  972    &   2   &  0.16   &   0.01   \\
BD$+$29$^{\circ}$3070\tablefootmark{d}  &  1283   &   63  &  0.15   &   0.01   \\
BD$-$7$^{\circ}$5977 \tablefootmark{e}  &  1314   &   8   &  0.15   &   0.01   \\
Bal\,8280003\tablefootmark{e}           &  1363   &   25  &  0.25   &   0.02   \\
\hline
\end{tabular}
\tablefoot{\tablefoottext{a}{\citet{Barlow12}}, \tablefoottext{b}{\citet{Vos12}}, \tablefoottext{c}{\citet{Deca2015}}, \tablefoottext{d}{\citet{Vos13}}, \tablefoottext{e}{\citet{Vos14}}}
\end{table}

Of the systems with the shortest periods, two have a circular orbit (PG\,1104$+$243, PG\,1701$+$359) and one has a slightly eccentric orbit (PG\,1018$+$243). Models should thus account for both circular and eccentric systems. Furthermore, there is a clear trend detected in Fig.\,\ref{fig_observed_period_ecc} in which the eccentricity increases with increasing orbital period, ranging from $e = 0.05$ at P = $\sim$750 d to e = 0.25 at P = $\sim$ 1350 d.

\section{MESA}\label{sec_mesa}
Modules for Experiments in Stellar Astrophysics (MESA)\footnote{MESA is available on: \url{http://mesa.sourceforge.net/}} is an open-source state-of-the-art 1D stellar-evolution code, which amongst others, includes a binary module to compute evolutionary tracks of binary stars. The stellar-evolution modules of MESA are extensively described in the two instrument papers: \citet{Paxton2011, Paxton2013}. In this article we have used version 7211 of MESA. A reason to work with this evolution code is the availability as open source. Moreover, the code calculates stellar models of low-mass stars through the helium flash.

The binary module of MESA is under continuous development. We started from version 7211 and extended it with several physical processes necessary for this research. The most important of which are:
\begin{itemize}
 \item Accretion of the mass lost in stellar winds. To calculate the accretion rates, the Bondy-Hoyle formalism as described in \citet{Hurley2002} is used. 
 \item Tidally-enhanced mass loss. Wind mass loss can be enhanced by the tidal influence of the companion star. The Companion Reinforced Attrition Processes (CRAP) mechanism of \citet{Tout1988} is used to calculate this.
 \item Phase-dependent mass loss during RLOF or phase-dependent mass loss through stellar winds can possibly increase the eccentricity of the orbit. The formalism of \citet{Soker2000} and \citet{Eggleton2006} is used to determine the change in eccentricity due to mass lost to infinity and mass accreted by the companion.
 \item Circumbinary (CB) disks. Due to Lindblad resonances, CB disk-binary interactions can change the orbital period and eccentricity. The formalism of \citet{Artymowicz1994} and \citet{Lubow1996} is used to calculate the CB disk-binary interaction.
\end{itemize}
A list of the processes implemented in the binary module is given in appendix \ref{app_mesa}. We limited ourselves to listing the physical processes. For the exact software implementation, we refer the reader to the open-source code, and future instrument papers. Most of the binary additions summarised above, will be added to MESA in the future. Until then, a copy of the binary module used here will be made available\footnote{The binary module is available on: \url{http://ster.kuleuven.be/~jorisv/MESA_binary}}.

\begin{figure*}[!t]
\centering
\includegraphics{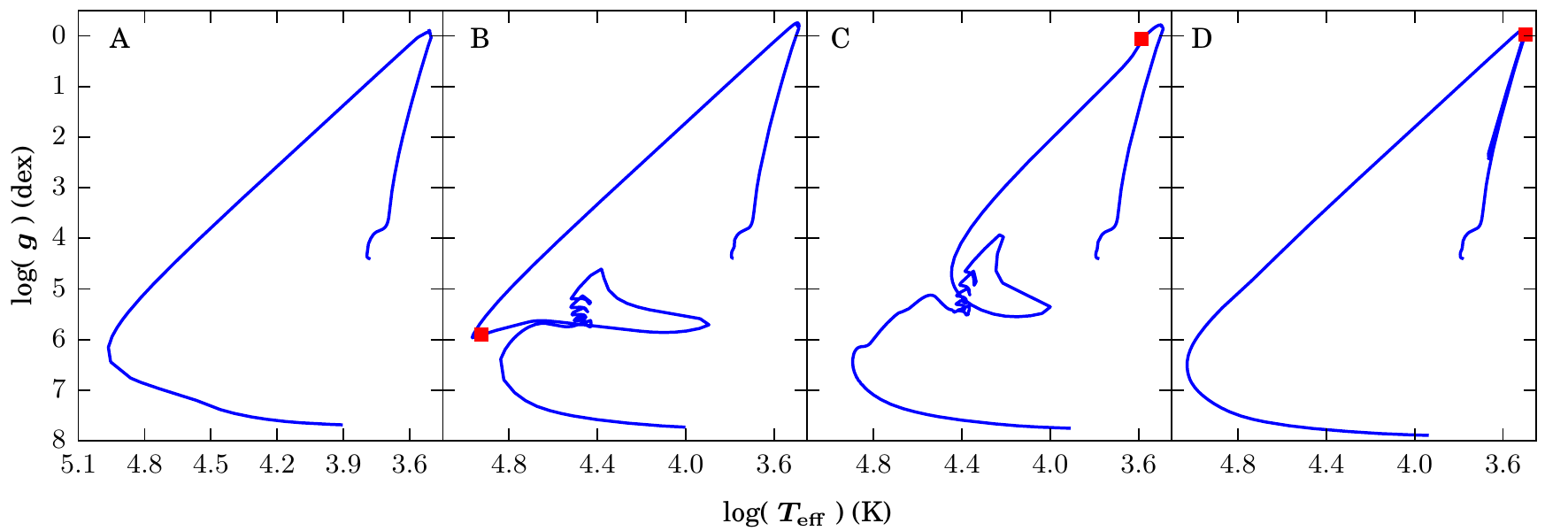}
\caption{HR diagram of the evolution of the sdB progenitor for different values of the initial period. The location of the He ignition is indicated with a red square. Panel A: P = 600 d, the donor star loses too much mass to ignite He and ends up as a cooling white dwarf (M$_d$ = 0.448 $M_{\odot}$). Panel B: P = 650 d, a late He flasher (M$_d$ = 0.456 $M_{\odot}$). Panel C: P = 750 d, an early He flasher (M$_d$ = 0.466 $M_{\odot}$). Panel D: P = 800 d, The core is too massive, and the donor ignites He on the tip of the RGB (M$_d$ = 0.929 $M_{\odot}$). See section \ref{sec_rlof} for discussion.}\label{fig_rlof_period_HR_diagram}
\end{figure*}

As an example of binary tracks leading to post-RGB evolution, the evolutionary tracks in the HR diagram for a late hot flasher and an early hot flasher are displayed in panels B and C of Fig. \ref{fig_rlof_period_HR_diagram}. 

In this article we will refer to the sdB progenitor as the mass-losing star or donor star, using the subscript d. The companion is also called the accretor, even when it won't accrete mass, using subscript a.

\subsection{Stellar input parameters}
As the main focus of this contribution is the study of binary-evolution processes, we will use standard parameters for the evolution of the individual stellar components. These include a standard atmospheric boundary condition at an optical depth of $\tau\,=\,2/3$, a mixing-length parameter of $\alpha_{\mathrm{MLT}}\,=\,2$ and default opacity tables ({\sc OPAL type i}). For the sdB progenitor we used an extended version of the standard nuclear networks to include all reactions for hydrogen and helium burning (\texttt{pp\_cno\_extras\_o18\_ne22.net}). The initial composition is  X = 0.68, Y = 0.30 and Z = 0.02. Furthermore we used a Reimers wind on the RGB with $\eta_{\mathrm{reimers}} = 0.7$ \citep{Reimers1975}, and a Bl\"{o}cker wind scheme on the post-EHB with $\eta_{\mathrm{blocker}} = 0.5$ \citep{Blocker1995}.
All models are calculated from the pre-main sequence till the WD-cooling curve, with the RGB taking the longest computation time. The sdBs resulting from these standard settings do not completely reproduce the observed effective temperature or surface gravity. This has been reported before, for example by \citet{Oestensen2012b, Oestensen2014}. We are aware of the discrepancy, but solving this is beyond the scope of this article.

\section{Modelling methodology}\label{sec_methodology}
The focus of this article is the effect of the different eccentricity-pumping mechanisms on the evolution. Three methods are investigated in the following sections. Tidally-enhanced wind mass-loss is considered separately from the other two mechanisms. The main argument is that tidally-enhance wind mass-loss increases the wind mass loss so that the donor star never fills its Roche lobe. In Sect.\,\ref{sec_rlof} the effects of firstly phase-dependent RLOF alone, and secondly phase-dependent RLOF in combination with a CB disk are described (Sect.\,\ref{sec_disk}). As the CB disk is created from mass lost during RLOF, the eccentricity-pumping effects of phase-dependent RLOF will also be present when the CB disk-binary interaction starts. We do not consider the CB disk-binary interactions separately, but assume the disk is formed by mass lost in the previous phase. 
For the phase-dependent RLOF models and the CB-disk models, a default model is used to show the evolution of several binary parameters. After which the effect of both initial binary parameters and method dependent parameters are discussed. The complete range in period an eccentricity that can be covered using any of the eccentricity-pumping mechanisms is discussed separately. 

In the case of phase-dependent RLOF, and the CB disk-binary interaction models, tidal forces will have circularised the orbit before the onset of RLOF. However, if the eccentricity is zero, the eccentricity-pumping effect of phase-dependent mass loss and CB disk-binary interactions will be zero as well. Therefore, we impose a lower limit on the eccentricity of $0.001$. This lower limit is chosen small enough to be reasonable. This minimum eccentricity is applied in all models, also those with only tidally-enhanced wind mass-loss. Thus if we refer to circularised models, this effectively means models that reached the minimum eccentricity of e = $0.001$.

For each eccentricity-pumping method a substantial number of models with varying initial parameters and process-depending parameters are calculated. The process-dependent parameters and their ranges are discussed in the separate sections. The initial binary parameters (period, eccentricity and mass of both components) are determined based on the type of systems that we want to produce. We are focusing on the long period sdB binaries, thus initial orbital periods vary between 500 and 900 days. The sdB progenitor mass varies between 1.0 and 1.5 $M_{\odot}$. While the initial companion mass varies between 0.8 and 1.45 $M_{\odot}$, and is always lower than the sdB-progenitor mass. Apart for the tidally-enhanced wind mass-loss models, all models circularise completely before the eccentricity-pumping mechanisms become effective. Changing the initial eccentricity for these models then only affects the initial orbital momentum of the binary, thus changing $e_i$ has a similar effect as changing the initial period. 

Only a small subsection of this parameter space will result in binaries containing an sdB component. In the discussion of the effect of the process-depending parameters, only the models containing an sdB component are used. The selection criteria for sdBs are based on the stellar mass at He ignition, which has to be below 0.55 $M_{\odot}$. And the absence of a hydrogen burning shell. In our models, there is a clear mass gap between the sdB models, and the models that ignite He on the RGB. The latter having final masses of M$_\dsub\,\gtrsim$\,0.7\,$M_{\odot}$, while the sdB models have final masses of $M_\dsub\,\lesssim$\,0.49\,$M_{\odot}$. Models that don't ignite He are obviously not sdBs either.

\section{Tidally-enhanced wind mass-loss}\label{sec_crap}
\citet{Siess2014} has modelled the long-period eccentric system IP Eri by using tidally-enhanced wind mass-loss in combination with the eccentricity-pumping effect of phase-dependent mass loss. By increasing the wind mass loss, the radius of the donor can be kept small in comparison to its Roche-lobe, thus the tidal forces that circularise the system are weaker. By using the mechanism of \citet{Tout1988}, the wind mass loss depends on the orbital phase, with it being stronger at periastron than apastron. This difference between the mass lost at periastron and apastron can increase the orbital eccentricity, depending on the mass ratio of the system and the fraction of the wind mass loss that is accreted by the companion. 

The enhanced-wind-mass-loss model depends on only one parameter, $B_{\mathrm{wind}}$, which \citet{Tout1988} estimate at $B_{\mathrm{wind}} = 10^4$ to explain the mass inversion in the pre-RLOF system Z-Her. The wind-mass-loss rate in function of the orbital phase $\theta$ is:
\begin{equation}
 \dot{M}_{\mathrm{wind}}(\theta) = \dot{M}_{\mathrm{Reimers}} \cdot \left\{ 1 + B_{\mathrm{wind}} \cdot \min\left[ \left(\frac{R}{R_{\rm L}(\theta)}\right)^6, \frac{1}{2^6} \right] \right\},
\end{equation}
where $R$ is the stellar radius and $R_{\rm L}(\theta)$ is the Roche-lobe radius at phase $\theta$. A detailed description of the wind mass loss and accretion fractions in MESA is given in appendix \ref{app_wind_ml_mesa}.
The strength of the eccentricity pumping will depend on the mass-loss rate, and the fraction that is accreted by the companion. We follow the model proposed by \citet{Soker2000} to calculate the change in eccentricity:
\begin{equation}
 \dot{e}_{\mathrm{ml}} = \int_{\theta} \left[ \frac{|\dot{M}_{\infty}(\theta)|}{M_\dsub + M_\asub} + 2 |\dot{M}_{\mathrm{acc}}(\theta)| \left( \frac{1}{M_\dsub} - \frac{1}{M_\asub} \right) \right]  ( e + \cos{\theta} )\,{\rm d}\theta,
\end{equation}
where $M_\dsub$ and $M_\asub$ are the donor and accretor mass, $\dot{M}_{\infty}$ is the mass lost at phase $\theta$ to infinity, $\dot{M}_{\mathrm{acc}}$ is the mass lost at phase $\theta$ accreted by the companion and $e$ is the eccentricity. Mass lost to infinity increases the eccentricity. Mass that is accreted, drives a change in eccentricity that is positive only if the donor mass is lower than the accretor mass. See appendix \ref{app_phase_dependent_ml_mesa} for a detailed description of this mechanism. 

\begin{figure}[!t]
\centering
\includegraphics{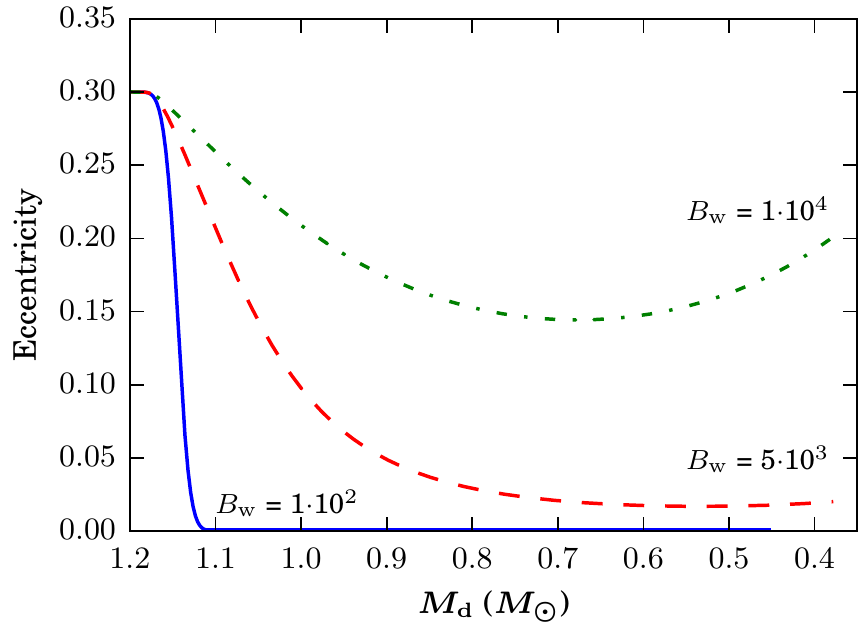}
\caption{The eccentricity in function of donor mass for models with  different values for wind-enhancement parameter $B_{\mathrm{wind}}$: 100 (solid blue), 5\,000 (dashed red) and 10\,000 (dot-dashed green). The initial period is 600 days, and the initial donor and companion masses are 1.2 + 0.8 $M_{\odot}$. For discussion see section \ref{sec_crap}. }\label{fig_crap_bwind}
\end{figure}

In Fig. \ref{fig_crap_bwind}, the eccentricity as a function of the mass of the donor star is plotted for a binary model with initial period of 600 days, a donor mass and companion mass of 1.2 and 0.8 $M_{\odot}$, and different values of B$_{\mathrm{wind}}$: 100, 5\,000 and 10\,000. As can be seen, this method can result in a significant final eccentricity of the orbit after the RGB evolution of the primary, if the enhancement parameter is sufficiently large. For the system shown in Fig.\,\ref{fig_crap_bwind}, the circularisation is overcome when $B_{\mathrm{wind}}$ $\gtrsim$ 5000, while an sdB star is only formed when 10 $\lesssim$ $B_{\mathrm{wind}}$ $\lesssim$ 100. 
The maximum wind-mass-loss rates in these systems are: $\dot{M}_{\mathrm{wind, max}}$ = $10^{-6.5}$, $10^{-6.2}$, $10^{-6.0}\,M_{\odot}$ yr$^{-1}$ for the systems with respectively $B_{\mathrm{wind}}$ = 100, 5\,000 and 10\,000. For comparison, the wind-mass-loss rate for the same system without enhancement would be $\dot{M}_{\mathrm{wind, max}} = 10^{-6.8}\,M_{\odot}$~yr$^{-1}$.

When comparing all calculated models, we find that the amount of mass that the donor needs to lose to maintain an eccentric orbit is so high that the final core mass is too low for He ignition. The donor star will then end its life as a cooling He-WD on an eccentric orbit. To reach a final donor mass high enough to ignite He, the mass loss has to be lower so that the system completely circularises on the RGB. Therefore, the sdB binaries formed in this channel are all circularised. 
The final mass of the donor changes with the initial parameters of the system, like orbital period, donor and accretor mass, and wind accretion fraction, but none of the combinations of initial binary parameters resulted in an eccentric system containing an sdB companion. \citet{Siess2014} gives a good overview of the effect of the initial binary parameters and the wind-enhancement parameter on the final parameters of the binary system. 

We conclude that in the parameter regime considered here, the progenitors of sdBs do not evolve with a strong enhanced mass loss on the RGB. To form an sdB by tidally-enhanced wind mass-loss, the enhancement of the wind needs to be small, so essentially no eccentricity pumping occurs. Furthermore, if there is a significant enhanced wind mass-loss, the sdB progenitor will lose its hydrogen envelope in a stellar wind, and will not undergo RLOF during its later evolution.

\section{phase-dependent RLOF}\label{sec_rlof}
When there is no strong enhanced wind mass loss during the RGB, the sdB progenitor will eventually fill its Roche-lobe and start RLOF. If this mass loss happens on a slightly eccentric orbit, the mass-loss rate will not be constant over the orbit, and the mass loss can have an eccentricity-pumping effect. The strength of the eccentricity pumping will depend on the mass-loss rate, and the fraction that is accreted by the companion, in the same way as in the tidally-enhanced-mass-loss mechanism.

\subsection{Model and input parameters}
In this section we will describe behaviour of the mass lost from the donor star in our study of phase-dependent RLOF. This mass can be accreted by the companion star, or is lost from the system to infinity. To describe the mass lost from the system to infinity, the method of \citet{Tauris2006} is used. A detailed description of that mechanism is given in appendix \ref{app_rlof_mesa}.
The total mass loss from the system is subdivided in three fractions: from around the donor star ($\alpha$), which carries the angular momentum of the donor; mass transferred to the vicinity of the companion through the inner Lagrange point, and lost from around the companion as a fast wind ($\beta$).  This lost mass carries the angular momentum of the companion star; And through the outer Lagrange point ($\delta$) which is modelled as a circumbinary toroid with radius 1.25 times the binary separation \citep[$R_{\mathrm{toroid}} = \gamma^2 a$, $\gamma$ = 1.12, based on][]{Pennington1985}.
The actual fraction of the mass loss accreted by the companion is defined by the aforementioned fractions as: $\epsilon = 1-\alpha - \beta - \delta$.

The default system has an initial period of 700 days, initial eccentricity of 0.3, an sdB progenitor mass of 1.2 $M_{\odot}$ and a companion mass of 1.0 $M_{\odot}$. The companion star will be on the main sequence during the relevant part of the evolution, thus the radiative dissipation mechanism for the tidal energy is assumed. The sdB progenitor will be on the RGB when the tidal forces are strongest, and the dissipation mechanism is that of a convective star. Dissipation becomes radiative when the envelope is completely lost. A maximum value for the RLOF mass-loss rate is set at $10^{-2}\,M_{\odot}$ yr$^{-1}$, the mass-loss fractions are $\alpha$ = $\beta$ = 0.35, $\delta$ = 0.30 and the location of the outer Lagrange point is $R_{\rm L2} = 1.25\,a$, thus $\gamma$ = 1.12. The accretion fraction onto the companion star is zero in this default model. These parameters are summarised in Table \ref{tb_standard_model_parameters}. Only the parameters mentioned in the text are changed.

As it is not known when exactly a common envelope would start to form we assume that RLOF is stable in our models. To this end, we have capped the mass-loss rate during RLOF at $10^{-2}\,M_{\odot}$ yr$^{-1}$ and applied an ad-hoc upper limit on the Roche-lobe overfilling of max($R / R_{\rm L}$) = 1.25, meaning that models in which the donor star radius exceeds 1.25 $\cdot$ $R_{\rm L}$ are discarded. The latter limit is also imposed as our physical model breaks down at high Roche-lobe overfilling. It is still possible that in these circumstances a CE would form during RLOF. However, \cite{Nelemans2000} argues that if a CE would form during unstable RLOF in systems with a long orbital period, the CE would co-rotate with the binary, and there would be no friction between the binary and the CE, and consequently no spiral-in phase. Furthermore, based on BPS studies, \cite{Clausen2012} finds that the existence of long-period sdB systems with MS components indicates that RLOF at the tip of the RGB is stable, or 
otherwise, that a CE phase without spiral-in does exist.

\begin{table}
\caption{Standard parameters of the binary models. For the models focusing only on phase-dependent RLOF, the disk mass is set to 0. The symbol column refers to the symbols used in the figures. }\label{tb_standard_model_parameters}
\centering
\begin{tabular}{lrr}
\hline\hline
\noalign{\smallskip}
Parameter	&	Symbol	&	Value\\\hline
\noalign{\smallskip}
sdB progenitor mass ($M_{\odot}$)	&	$M_\dsub$	&	1.2 \\
companion mass ($M_{\odot}$)		&	$M_\asub$	&	1.0 \\
period (d)				&	$P$	&	700 \\
eccentricity				&	$e$	&	0.3 \\
minimum eccentricity			&	/	&	0.001 \\
\noalign{\smallskip}
\multicolumn{3}{c}{Tidal forces}\\
dissipation type sdB		&	/	&	Convective \\
dissipation type companion	&	/	&	Radiative \\
\noalign{\smallskip}
\multicolumn{3}{c}{Mass loss}\\
maximum $\dot{M}$ ($M_{\odot}$ yr$^{-1}$)&	/	& 10$^{-2}$\\
fraction lost around companion	&	$\alpha$	&	0.35 \\
fraction lost around donor	&	$\beta$		&	0.35 \\
fraction lost through L2 	&	$\delta$	&	0.30 \\
location of L2			&	$\gamma$	&	1.12 \\
accreted fraction\tablefootmark{a} &	$\epsilon$ 	&	0 \\
\noalign{\smallskip}
\multicolumn{3}{c}{CB disk}\\
maximum mass ($M_{\odot}$)&	$M_{\mathrm{disk}}$	&	0.01 \\
life time (yr)		&	$\tau$			&	10$^5$ \\
viscosity		&	$\alpha_{\rm D}$	&	0.01 \\
mass distribution	&	$\sigma(r)$		&	r$^{-1}$ \\
\hline
\end{tabular}
\tablefoot{\tablefoottext{a}{The accreted fraction is defined by the other mass loss fractions as: $\epsilon = 1 - \alpha - \beta - \delta$.}}
\end{table}

\subsection{Parameter study}

\subsubsection{Eccentricity evolution}\label{sec_rlof_eccentricity_evolution}
\begin{figure}[!t]
\centering
\includegraphics{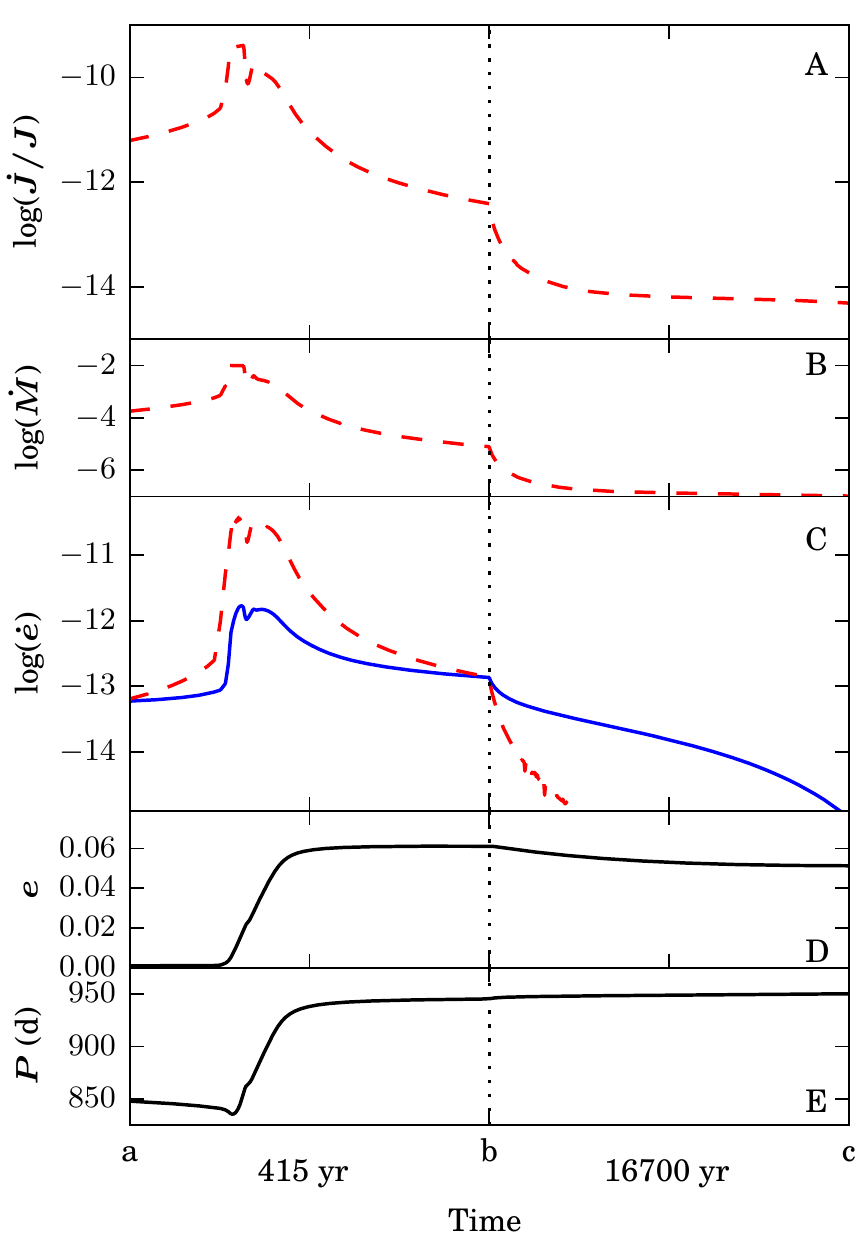}
\caption{Time evolution of several binary properties during the RLOF phase. Three different events are indicated on the X-axes, 
a: $\dot{e}_{\mathrm{ml}}$ > $\dot{e}_{\mathrm{tidal}}$, 
b: $\dot{e}_{\mathrm{ml}}$ < $\dot{e}_{\mathrm{tidal}}$ and 
c: $\log(\dot{e}_{\mathrm{tidal}})$ < -15. 
The time scale differs between the phases, but is linear withing each phase. The duration of each phase is shown under the figure. 
Panel A: the change in angular momentum. 
Panel B: mass loss through RLOF. 
Panel C: the tidal forces (blue) and eccentricity pumping due to mass loss (red). 
Panel D: the orbital eccentricity. 
Panel E: the orbital period. 
See section \ref{sec_rlof_eccentricity_evolution} for discussion.}\label{fig_rlof_standard_model}
\end{figure}

The evolution of the default model is plotted in Fig. \ref{fig_rlof_standard_model}. There are three events indicated on the figure, 
a: $\dot{e}_{\mathrm{ml}}$ > $\dot{e}_{\mathrm{tidal}}$, 
b: $\dot{e}_{\mathrm{ml}}$ < $\dot{e}_{\mathrm{tidal}}$ and 
c: $\log(\dot{e}_{\mathrm{tidal}})$ < -15. 
Where $\dot{e}_{\mathrm{ml}}$ is the eccentricity pumping of the mass loss in phase-dependent RLOF and $\dot{e}_{\mathrm{tidal}}$ is the circularisation due to stellar tides. These events indicate two phases in the period-eccentricity evolution of the binary. The timescale between these phases differs on the figure, and within one phase the time scale is linear. Underneath the figure the duration of each phase is given.
\begin{description}
 \item[Phase a--b:] The sdB progenitor is close to completely filling its Roche-lobe ($R / R_{\rm L}$ = 0.97), and the mass loss rate reached $10^{-3.7}\,M_{\odot}$ yr$^{-1}$, at which point the eccentricity-pumping forces are more effective than the tidal forces. Due to the increasing mass-loss rate, the eccentricity pumping stays stronger than the tidal forces, even though the latter increase due to further Roche-lobe overfilling ($\max R / R_{\rm L}$ = 1.09). The eccentricity starts increasing, and reaches a value of 0.061 by the end of this phase, an increase by a factor 60 compared to the assumed minimum eccentricity. During the mass-loss phase the period increases as well, from 847 to 945 days. When the star starts shrinking again, the mass-loss rate decreases as well as the Roche-lobe-filling factor, until event `b', when the tidal forces again become higher than the eccentricity pumping ($\dot{M} = 10^{-5.1}\,M_{\odot}$~yr$^{-1}$, $R / R_{\rm L}$ = 0.85). This phase of strong mass loss only takes 415 
years. 

 \item[Phase b--c:] When the RLOF diminishes, the tidal forces take over, and the binary starts circularising again for another 16700 years. During this phase the sdB progenitor is contracting, thus the tidal forces are weakening. The eccentricity diminishes from 0.061 to 0.051, while the period slightly increases to 950 days. For this system, the eccentricity increased by roughly a factor 50. The evolution is plotted to the point where $\dot{e}_{\mathrm{tidal}} < 10^{-15}$, after which there is no more significant change in eccentricity or period.
\end{description}

\subsubsection{Effect of initial period and companion mass}\label{sec_rlof_period_mass_effect}
The effect of the initial period and companion mass is plotted in Fig. \ref{fig_rlof_par_period_ecc}-A. With increasing initial period, the final period of the sdB binary will increase as well, while at the same time the eccentricity decreases. A similar effect is visible when increasing the companion mass. The closer the companion mass is to the donor mass of 1.2 $M_{\odot}$, the larger the final orbital period, and the lower the eccentricity. We first discuss the effect of the initial binary parameters, as they clearly illustrate the effect of the orbital period on the eccentricity pumping force that is also essential in the discussion of the process dependent parameters. 

\begin{figure}[!t]
\centering
\includegraphics{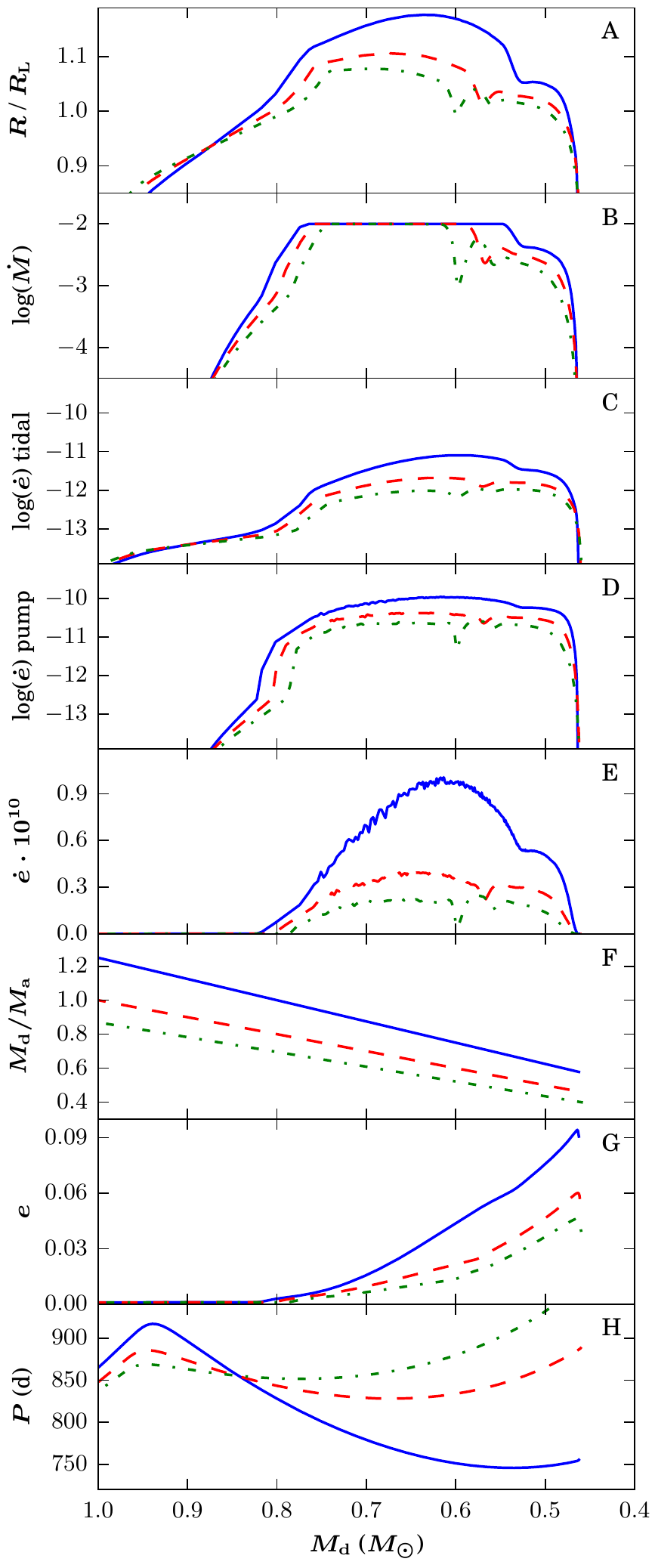}
\caption{Comparison of three models with different values for the companion mass:  0.8 $M_{\odot}$ (solid blue), 1.0 $M_{\odot}$ (dashed red) and 1.15 $M_{\odot}$ (dot-dashed green). 
Panel A: the Roche-lobe overfilling factor.  
Panel B: the mass loss rate during RLOF in $M_{\odot}$~yr$^{-1}$. 
Panel C: the tidal forces in log(s$^{-1}$). 
Panel D: the eccentricity pumping forces in log(s$^{-1}$). 
Panel E: net change in eccentricity per second. 
Panel F: mass ratio. Panel G: the eccentricity. 
Panel H: the orbital period in days. 
See section \ref{sec_rlof} for discussion. }\label{fig_rlof_companion_mass_history}
\vspace{-0.7cm}
\end{figure}

To explain these effects, the evolution of several parameters for models with different companion masses of 0.8, 1.0 and 1.15 $M_{\odot}$ are shown in Fig. \ref{fig_rlof_companion_mass_history}. These parameters are plotted in function of the donor mass instead of time, so that the different models can be more easily compared. The connection between eccentricity and orbital period is found in the change in mass-loss rate during RLOF. If the period of the system increases, the Roche-lobe overfilling is lower, varying from $R/R_{\rm L}$ < 1.16 for the $M_\asub$ = 0.8 $M_{\odot}$ system to $R/R_{\rm L}$ < 1.09 and 1.07 for $M_\asub$ = 1.0 and 1.15 $M_{\odot}$ respectively. This lower overfilling of the Roche lobe leads to a lower mass-loss rate and a shorter total time during which mass is lost at the maximum rate. 
For the models described here, the maximum mass-loss rate is the same, but the time during which it is sustained differs from roughly 25 years for the 0.8 $M_{\odot}$ companion to 20 years for the 1.0 $M_{\odot}$ companion and 15 years for the 1.15 $M_{\odot}$ companion. The difference in time scale seems small, but is significant when the mass-loss rate is 10$^{-2}$ $M_{\odot}$ yr$^{-1}$. This diminishes the total eccentricity pumping force, even though the time during which it overpowers the tidal forces is longer for the 1.15 $M_{\odot}$ model ($\sim$ 380 yr) than for the 0.8 $M_{\odot}$ model ($\sim$ 200 yr). Due to the lower Roche-lobe overfilling, the tidal forces will also decrease with increasing orbital period, with a factor 10 difference between the 0.8 $M_{\odot}$ model and the 1.15 $M_{\odot}$ model. Even so, there is a net effect of lower eccentricity enhancement for the model with the heaviest companion. 

In Fig. \ref{fig_rlof_par_period_ecc} only the period and mass range of models that resulted in an sdB binary are shown. For a system with a specified donor and accretor mass, the effect of the initial period on the final mass is important. If the orbital period is too short, the mass loss will be too strong, and the donor star will lose too much mass to ignite helium, ending up on the He-WD cooling track. The lower limit on helium ignition in MESA is around 0.45 $M_{\odot}$, slightly depending on composition. If the initial orbital period is too high, the mass-loss rate during RLOF will be too low, and the donor will ignite helium on the RGB and hence no sdB is created. In Fig. \ref{fig_rlof_period_HR_diagram} the HR diagram of four models with a donor and companion mass of 1.2 + 1.0 $M_{\odot}$ and initial periods of 600, 650, 750 and 800 days are shown. 
The system with the lowest orbital period (panel A) ends up with a donor star mass of 0.448 $M_{\odot}$, this is just under the He-ignition limit, and the donor star ends its evolution as a He white dwarf. The system with the 650 days initial period (panel B) ends with a donor star of 0.456 $M_{\odot}$ and ignites helium on the He-WD cooling track, which is called a late hot flasher. With an even higher period of 750 days, the final donor mass is 0.465 $M_{\odot}$, and the donor ignites helium shortly after departing of the RGB, during the evolution at constant luminosity, an early hot flasher. If the period is increased further, to 800 days, the donor will be too massive, and will ignite helium on the RGB when it still has a total mass of 0.929 $M_{\odot}$, after which mass loss is stopped until the donor finishes core-He burning and enters the AGB phase.

\subsubsection{Effect of mass-loss fractions and accretion}\label{sec_rlof_massloss_effect}
\begin{figure*}[!t]
\centering
\includegraphics{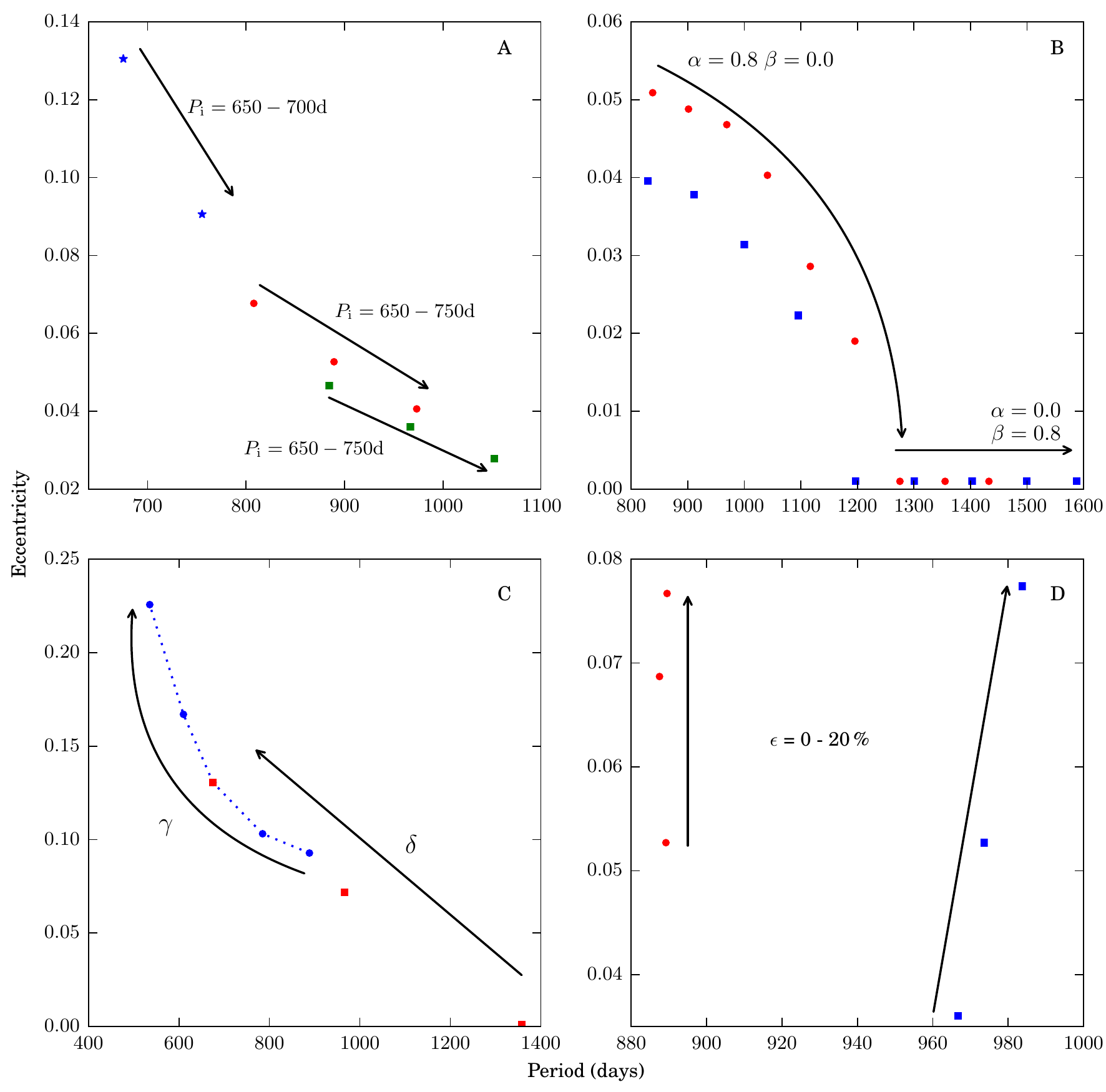}
\caption{The effect of the initial and mass loss parameters of the RLOF models on the final period and eccentricity of the orbit. 
Panel A: Models with different initial periods (indicated by the black arrows) and different companion masses: $M_\asub$ = 0.8 (blue stars), $M_\asub$ = 1.0 (red dots) and $M_\asub$ = 1.15 (green squares).
Panel B: Models with different mass loss fractions $\alpha$ and $\beta$, for two companion masses: 1.0 $M_{\odot}$ (red circles) and 1.15 $M_{\odot}$ (blue squares). The black arrow indicates decreasing $\alpha$ and increasing $\beta$, $\delta$ is constant at 0.2. The sum of the mass loss fractions is unity for each model. 
Panel C: Models with different mass loss fractions $\delta$: 0.1, 0.2 and 0.3, with $\gamma = 1.12$ (red squares) and different values for $\gamma$: 0.9, 1.0, 1.2 and 1.3 with $\delta = 0.3$ (blue dots). These models don't have accretion, thus $\alpha = \beta = (1 - \delta)/2$.
Panel D: The effect of the accreted fraction for two companion masses and accretion fractions of 0\%, 10\% and 20\%. $M_\asub$ = 1.0 $M_{\odot}$ (red circles) and $M_\asub$ = 1.15 $M_{\odot}$ (blue squares). 
See Sects. \ref{sec_rlof_period_mass_effect} and \ref{sec_rlof_massloss_effect} for discussion.}\label{fig_rlof_par_period_ecc}
\end{figure*}

The different ways to lose mass to infinity will change the final orbital parameters of the system. These parameters mainly influence the amount of angular momentum that is removed from the system with the lost mass. The effect of these mass-loss fractions and the location of the outer Lagrange point is shown in figures \ref{fig_rlof_par_period_ecc}-B and \ref{fig_rlof_par_period_ecc}-C. 
In Fig. \ref{fig_rlof_par_period_ecc}-B, models for two different companion masses 1.0 and 1.15 $M_{\odot}$ at a constant $\delta$ = 0.2 are plotted, while changing the fractions of mass lost around the donor ($\alpha$) and companion ($\beta$). There is no accretion in these models ($\alpha + \beta + \delta = 1$). When most mass is lost from around the donor the resulting period will be lower than when the mass is lost from around the companion star. 
This is easily explained by eq. \ref{eq_mdot_jdot_tauris}. The mass that is lost from around the donor star will thus carry more angular momentum with it, which will result in a shorter orbital period. This change in period will influence the change in eccentricity by altering the mass-loss rates during RLOF, similar as was explained in Sect.\ref{sec_rlof_period_mass_effect}. From a certain threshold period that depends on the initial and mass-loss parameters, the eccentricity pumping is smaller than the tidal forces, and the orbits stay circularised. The effect of these two mass-loss parameters on the period is large, on the order of several hundred days. By changing from most mass lost around the donor to most mass lost around the companion, the final period can double.

By increasing the fraction of mass that is lost through the outer Lagrange point ($\delta$), more angular momentum is lost than if that mass was lost from around either of the binary components. The decrease in final period with increasing $\delta$ then follows directly from eq. \ref{eq_mdot_jdot_tauris}. Increasing the size of the circumbinary toroid representing the location of the outer Lagrange point ($\gamma$) has the same effect, for exactly the same reason. Fig. \ref{fig_rlof_par_period_ecc}-C shows the effect of changing $\delta$ and $\gamma$. The models with varying $\delta$ have no accretion, thus $\alpha = \beta = (1 - \delta)/2$. The increase of the eccentricity with increasing $\delta$ and $\gamma$ is directly related to the decrease of the orbital period, and thus a higher mass-loss rate. 
The two models on Fig. \ref{fig_rlof_par_period_ecc}-C with $\gamma$ values of 1.2 and 1.3 result in binaries with very high Roche-lobe-overfilling factors ($R / R_{\rm L} > 1.6$). Note that our description breaks down at high Roche-lobe-overfilling values, and the models in Fig.\,\ref{fig_rlof_par_period_ecc}-C with high values for $\gamma$ suffer from extrapolation. The effect of $\gamma$ is only given as an indication, as we keep it at 1.12 to represent the expected location of the outer Lagrange point. By changing $\delta$ by only a small amount, the final period can again change drastically, on the order of hundreds of days. Similar to the $\alpha$ and $\beta$ fractions, there is a certain threshold for $\delta$ under which the orbit will stay circularised. The exact threshold value depends on the other parameters.

The effect of the mass-loss fraction accreted onto the companion ($\epsilon$) is shown in panel D of Fig. \ref{fig_rlof_par_period_ecc}. By increasing the accreted fraction, the eccentricity increases while the orbital period stays more or less constant. By accreting a certain fraction of the mass loss, the size of the Roche-lobes will differ between the models. Higher accretion leads to slightly higher Roche-lobe overfilling, and a slightly higher eccentricity pumping. The period during which the eccentricity-pumping forces are stronger than the tidal forces also increases with increasing accretion rates. By increasing the accretion, the final eccentricity can almost be doubled, while the period remains constant.

\section{Circumbinary Disks}\label{sec_disk}
The models with phase-dependent RLOF can indeed explain a certain part of the period-eccentricity diagram, but have problems in the high-period high-eccentricity range, and cannot reproduce the circular systems. Circumbinary disks (CB disks) could potentially explain the high-period, high-eccentricity systems as they add extra eccentricity-pumping forces on top of those from phase-dependent RLOF. 

\subsection{Model and input parameters}
CB disks can form around binaries during the Roche-lobe overflow phase, if part of the mass can leave the system through the outer Lagrange points and form a Keplerian disk around the binary. The CB disk--binary resonant and non-resonant interactions have been described by \citet{Goldreich1979} and \citet{Artymowicz1994}, by using a linear-perturbation theory. The effect of the CB disk--binary resonances on the orbital parameters has been the subject of many studies. In MESA we follow the approach outlined by \citet{Artymowicz1994} and \citet{Lubow1996}. 

The effect of the disk on the binary separation is given by:
\begin{equation}
 \frac{\dot{a}}{a} = -\frac{2l}{m} \cdot \frac{J_{\rm D}}{J_{\rm B}} \cdot \frac{1}{\tau_{\rm v}},
\end{equation}
where $J_{\rm D}$ is the angular momentum of the disk and $J_{\rm B}$ the orbital angular momentum of the binary. $l$ and $m$ are integer numbers indicating the resonance with the strongest contribution and $\tau_{\rm v}$ is the viscous evolution timescale (where $\tau_{\rm v} \sim \alpha_{\rm D}^{-1}$, see eq.\,\ref{eq_disk_tau_visc}). The change in eccentricity depends on the change in binary separation as:
\begin{equation}
 \dot{e}_{\mathrm{disk}} = \dfrac{1-e^2}{e + \dfrac{\alpha}{100 e}} \left( \dfrac{l}{m} - \dfrac{1}{\sqrt{1 - e^2}} \right) \cdot \dfrac{\dot{a}}{a},
\end{equation}
for small eccentricities ($e < 0.2$), and decreases with $1/e$ for higher eccentricities. The derivation and implementation of these equations is given in appendix \ref{app_cb_disks_mesa}.

The CB disk is described by its maximum mass ($M_{\mathrm{disk}}$), the mass distribution, the inner and outer radius, relative thickness near the inner rim ($H/R$), the viscosity of the matter ($\alpha_{\rm D}$) and the total life time of the disk ($\tau$). In our model, the disk is formed by matter lost from the binary through the outer Lagrange point, and the disk itself loses mass at a rate determined by its life time. The mass in the disk is not constant, and only the maximum disk mass is a defined input parameter. The model of \citet{Artymowicz1994} is only valid for a thin disk. Thus, $H/R$ = 0.1 is assumed. The inner radius of the disk is determined based on smooth-particle-hydrodynamic (SPH) simulations, and the dust-condensation radius of the binary (see eq. \ref{eq_disk_rin_sph} -- \ref{eq_disk_rin}). Based on observations and the assumed surface density behaviour, the outer radius of the disk is fixed at 250 AU (see also appendix \ref{app_cb_disks_mesa}).

Maximum mass, life time, viscosity and distribution, are input parameters in the model. Based on observations of post-AGB disks \citep{Gielen2011, Hillen2014, Hillen2015}, following parameter ranges are assumed. The total disk mass can vary between $10^{-4}$ and $10^{-2}\,M_{\odot}$. Disk life times after the end of RLOF range from $10^4$ to $10^5$ years. The viscosity ranges from $\alpha_{\rm D} = 0.001 - 0.1$, with the more likely range being $\alpha_{\rm D} \leq 0.01$. The surface mass distribution decreases with increasing distance from the centre: $\sigma(r) \sim r^{D}$, where $-2 \leq D \leq -1$. In the default model given in Table\,\ref{tb_standard_model_parameters}, a maximum disk mass of $10^{-2}\,M_{\odot}$, a life time of $10^5$ years, a viscosity of 0.01 and a surface distribution of $\sigma(r) \sim r^{-1}$ are chosen.

\subsection{Parameter study}

\begin{figure}[!t]
  \centering
  \includegraphics{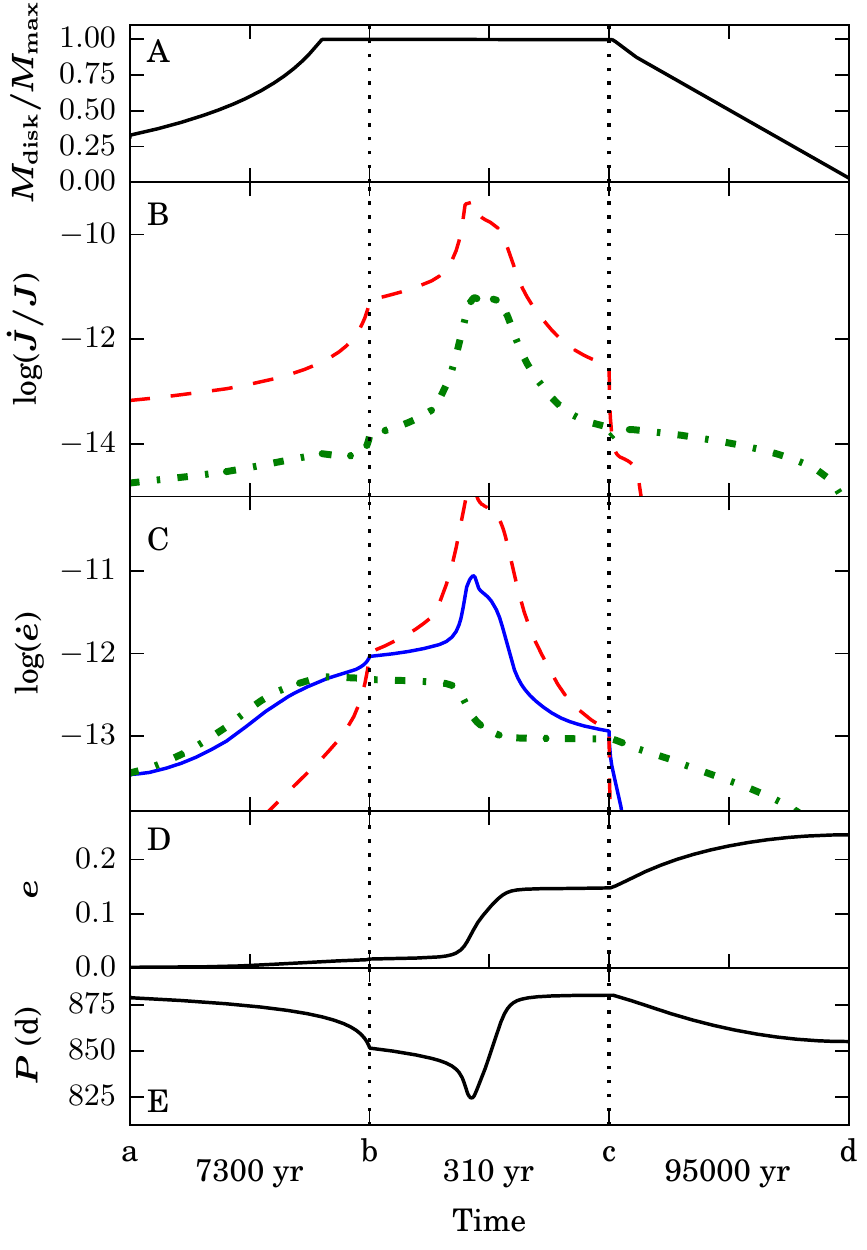}
  \caption{Time evolution of several binary properties during the life time of the CB disk. Four different events are indicated on the x-axes. 
  a: $\dot{e}_{\mathrm{disk}}$ > $\dot{e}_{\mathrm{tidal}}$, 
  b: $\dot{e}_{\mathrm{ml}}$ > $\dot{e}_{\mathrm{tidal}}$, 
  c: $\dot{e}_{\mathrm{ml}}$ < $\dot{e}_{\mathrm{tidal}}$ and 
  d: $\dot{e}_{\mathrm{disk}}$ < $\dot{e}_{\mathrm{tidal}}$. 
  The time scale differs between the phases, but is linear within each phase. The duration of each phase is shown under the figure. 
  Panel A: the mass in the CB disk as a percentage of the maximum CB disk mass (0.01 $M_{\odot}$). 
  Panel B: the change in angular momentum due to mass loss (red dashed line) and the CB disk - binary interaction (green dashed dotted line). 
  Panel C: the tidal forces (blue full line), eccentricity pumping due to mass loss (red dashed line) and eccentricity pumping through CB disk - binary interactions (green dashed dotted line). 
  Panel D: the eccentricity. 
  Panel E: the orbital period. 
  See section \ref{sec_disk} for discussion.}\label{fig_disk_standard_model}
\end{figure}

\subsubsection{Eccentricity evolution}
In Fig.\,\ref{fig_disk_standard_model} the evolution of the CB disk mass, the period and eccentricity, the change in angular momentum and the change in eccentricity are plotted in function of time. The model that is shown has the same parameters as the model displayed in Fig.\,\ref{fig_rlof_standard_model}, with the addition of a CB disk with a life time of $10^{5}$ yr. These parameters are also given in Table\,\ref{tb_standard_model_parameters}. Four different events are indicated on the figure. 
a: $\dot{e}_{\mathrm{disk}}$ > $\dot{e}_{\mathrm{tidal}}$, 
b: $\dot{e}_{\mathrm{ml}}$ > $\dot{e}_{\mathrm{tidal}}$, 
c: $\dot{e}_{\mathrm{ml}}$ < $\dot{e}_{\mathrm{tidal}}$ and 
d: $\dot{e}_{\mathrm{disk}}$ < $\dot{e}_{\mathrm{tidal}}$. 
These events define three different phases in the period-eccentricity evolution of the binary. On the figure, the time of each phase is linear, but the timescales between phases differ. The duration of each phase is plotted under the figure.
\begin{description}
 \item [Phase a--b:] Mass lost from atmospheric RLOF is filling the CB disk, and the disk - binary interactions are strong enough to overcome the tidal forces, thus eccentricity starts to increase. The mass in the disk continues to grow while the donor star continues to expand, eventually filling and overfilling its Roche-lobe. The tidal forces continue to increase as well, and when the CB disk reaches its maximum mass after roughly 6100 years, the tidal forces again overtake the disk-binary interaction. By this time the eccentricity of the system reached 0.016 and the period decreased from 878 to 852 days. The change in eccentricity is solely due to the disk-binary interactions, but the loss of angular momentum from the binary is caused mainly by the mass loss. During the remaining $\sim$1000 years when $\dot{e}_{\mathrm{disk}}$ < $\dot{e}_{\mathrm{tidal}}$, there is a negligible amount of circularisation. 
 \item [Phase b--c:] The disk reached its maximum mass, and is maintained by RLOF. Due to phase-dependent mass loss, the eccentricity continues to increase. What happens in this phase is very similar to what is shown in Fig\,\ref{fig_rlof_standard_model}. However, because the binary has a higher eccentricity than that in the model without a disk, the eccentricity pumping due to mass loss is stronger. This leads to an eccentricity of 0.146, a tenfold increase. The orbital period first decreases and then increases again, reaching 880 days. This whole phase lasts only 310 years. 
 \item [Phase c--d:] RLOF has ended, and the mass in the CB disk starts to decrease linearly. The CB disk-binary interactions continue to increase the eccentricity, while at the same time angular momentum is transported from the binary to the CB disk, thus decreasing the orbital period again. The effect of the CB disk-binary interactions diminishes due to a decrease in disk mass while simultaneously, the resonances that are responsible for the eccentricity pumping become less effective at higher eccentricities. When the disk is completely dissipated, the binary has an eccentricity of 0.245 and a period of 855 days. 
\end{description}
We note that the transport of angular momentum from the binary to the disk will continue to take place as long as the disk lives, contrary to the change in eccentricity that has an upper limit of e $\sim$ 0.7. However, no models we have tried reach this eccentricity limit. 

\begin{figure*}[!t]
\centering
\includegraphics{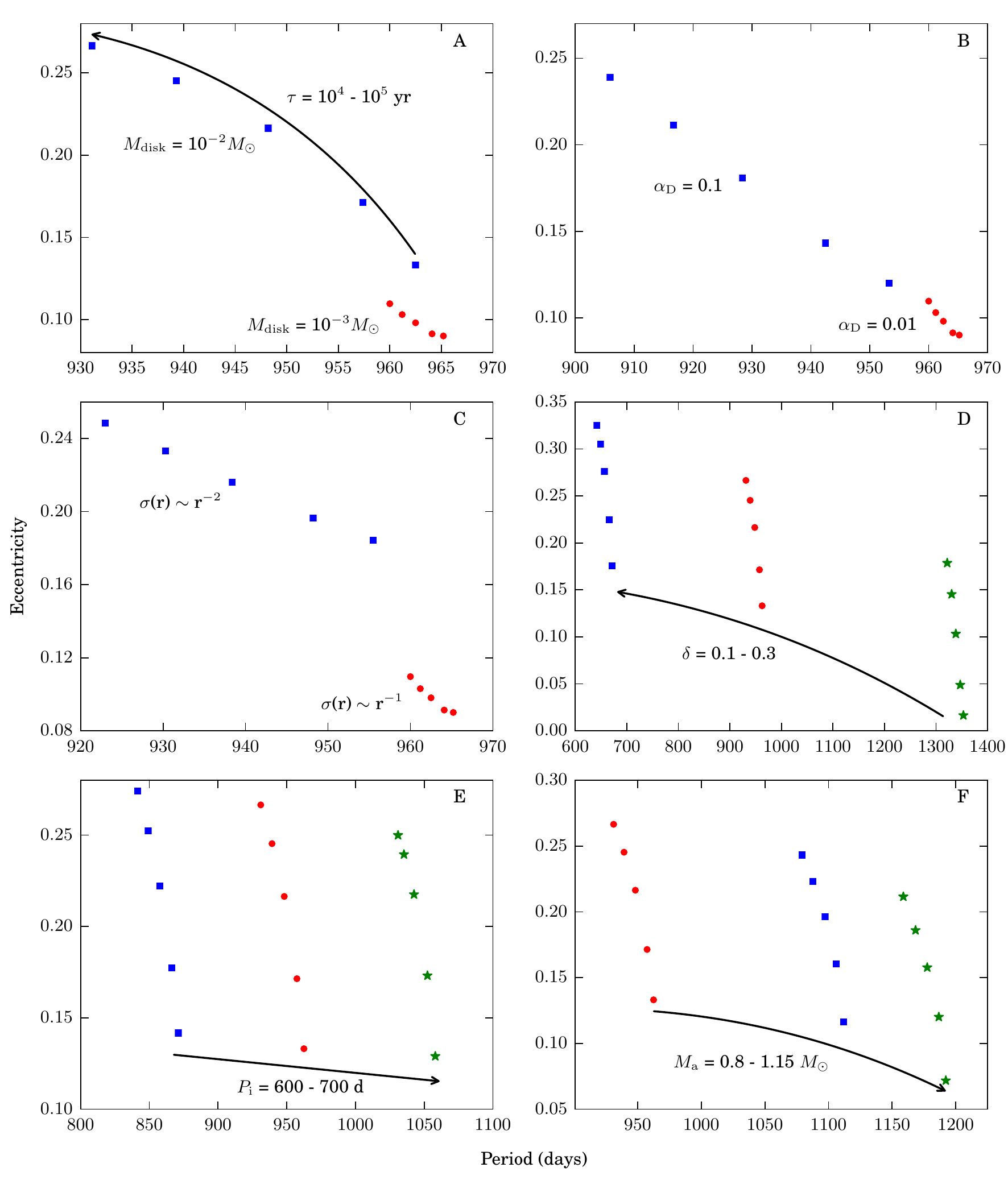}
\caption{The effect of the CB disk and binary parameters on the final period and eccentricity of the orbit. All models are plotted for 5 different disk life times $\tau$ = $1 \cdot 10^4$, $2.5 \cdot 10^4$, $5 \cdot 10^4$, $7.5 \cdot 10^4$ and $1 \cdot 10^5$ years.
Panel A: Models with two different maximum disk masses: 10$^{-2}$ $M_{\odot}$ (blue squares) and 10$^{-3}$ $M_{\odot}$ (red circles).
Panel B: Models with a different disk viscosities: $\alpha_{\rm D}$ = 0.1 (blue squares) and 0.01 (red circles).
Panel C: Models with two different disk surface mass distributions: $\sigma(r) \sim r^{-2}$ (blue squares) and $\sigma(r) \sim r^{-1}$ (red circles).
Panel D: Models with three different values for mass loss parameter $\delta$: 0.1 (blue squares), 0.2 (red circles) and 0.3 (green stars).
Panel E: Models with three different initial orbital periods: $P_{\rm i}$ = 600 d (blue squares), 650 d (red circles) and 700 d (green stars).
Panel F: Models for three different companion masses: $M_\asub$ = 0.8 $M_{\odot}$ (red circles), 1.0 $M_{\odot}$ (blue squares) and 1.15 $M_{\odot}$ (green stars).
See Sects. \ref{sec_disk_cb_effect} and \ref{sec_disk_binary_effect} for discussion.}\label{fig_disk_par_period_ecc}
\vspace{-0.5cm}
\end{figure*}

\subsubsection{Effect of disk properties}\label{sec_disk_cb_effect}
There are four parameters in the CB disk model that can influence the CB disk-binary interactions: the maximum mass in the disk, the life time of the disk, the viscosity parameter and the assumed distribution of the mass with radius. Next to the disk parameters, the rate at which mass is fed to the disk is determined by mass-loss fraction $\delta$.  The effect of these five parameters is shown in Fig.\,\ref{fig_disk_par_period_ecc} panels A, B, C and D. For each of maximum disk mass, viscosity, mass distribution and $\delta$, models with five different disk life times are shown. These life times can also be interpreted as the evolution in the period-eccentricity diagram after the end of RLOF. Based on the equations that govern the CB disk-binary interaction given in section \ref{app_cb_disks_mesa} the effect of each parameter can be easily explained. When the disk survives longer, the interaction with the binary will last longer, and the increase in eccentricity / decrease in period will be stronger. 

By increasing the maximum disk mass as shown in panel A, the final eccentricity will be higher, while the final period will be slightly lower. The mass-loss rates during RLOF are high enough that there is very little difference in time to fill a disk of 0.001 $M_{\odot}$ or 0.01 $M_{\odot}$. An increased disk mass will lead to a higher angular momentum of the disk, and, according to eq. \ref{eq_disk_dlna}, to a higher change in binary separation and eccentricity. 

The viscosity of the disk ($\alpha_{\mathrm{D}}$) shown in panel B, also has a linear effect on the change in binary separation and eccentricity (see eq. \ref{eq_disk_tau_visc}). A higher viscosity will lead to a higher final eccentricity and lower period. 

The effect of the mass distribution in the disk is plotted in panel C. The mass distribution will determine the mass fraction that resides close to the inner rim of the disk, where it has the strongest effect on the binary. Normally a radial mass distribution of $r^{-1}$ is chosen, which takes into account that the disk is not just flat, but also has a specific thickness that varies with the radius. A distribution of $r^{-2}$ assumes that the thickness of the disk is more constant with radius. Under that assumption, there will be more mass closer to the inner rim, and the CB disk-binary interactions will be stronger, again leading to a higher eccentricity and lower period. 

The rate at which mass is fed to the CB-disk is defined by mass-loss parameter $\delta$. The effect of changing this parameter is displayed in panel D. As the disk will reach its maximum mass earlier, the eccentricity-pumping effect of the disk will slightly increase with increasing $\delta$. However, the main effect of this parameter is a change in orbital period. As discussed in Sect.\,\ref{sec_rlof_massloss_effect}, increasing $\delta$ will decrease the orbital period because more angular momentum is lost. This results in higher eccentricities reached during the RLOF phase. 

\subsubsection{Effect of binary properties}\label{sec_disk_binary_effect}
In panels E and F of Fig.\,\ref{fig_disk_par_period_ecc}, the effect of several binary parameters is shown. These are also discussed in the Sect. \ref{sec_rlof_period_mass_effect} and are shown here, to indicate how strong their influence on the distribution in the period-eccentricity plane can be. It is especially the range in period space that is altered by changing initial parameters such as companion mass and initial period. Where the disk-binary interactions have an influence of roughly five percent on the final period, changing the initial binary parameters can result in a change of several 100 days. The main influence of the CB disk is then in reaching higher eccentricities. Where the phase-dependent RLOF models could reach eccentricities of maximum 0.15, the models containing a CB disk can reach eccentricities of maximum 0.34.

\section{Period-eccentricity distribution}\label{sec_period_ecc_distribution}
\begin{figure}[!t]
  \centering
  \includegraphics{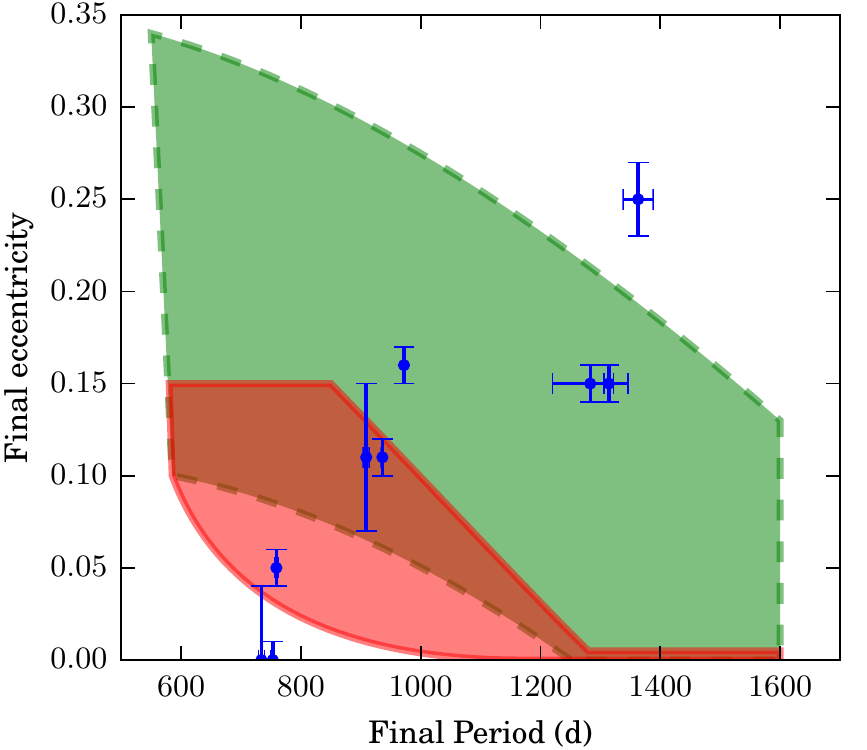}
  \caption{The approximate region of the period-eccentricity diagram that can be explained by the RLOF models (red shaded region with solid border) and models containing a CB disk (green shaded region with dashed border). The observed systems are plotted in blue dots.}\label{fig_period_ecc_model_coverage}
\end{figure}

By varying the initial binary parameters described in Sect.\,\ref{sec_methodology} and the model-dependent parameters described in Sects.\,\ref{sec_rlof} and \ref{sec_disk}, a significant area of the period-eccentricity diagram can be covered. In Fig. \ref{fig_period_ecc_model_coverage}, both the region covered by models with only phase-dependent RLOF (red shade with solid border) and that covered by models including a CB disk (green shade with dashed border) is plotted on top of the observed systems (blue circles). The parameter ranges used to obtain this distribution are given in Table\,\ref{tb_parameter_ranges}. As the models with only tidally-enhanced wind mass-loss cannot produce eccentric sdB binaries, they are not shown here. We see that by including eccentricity-pumping processes during the evolution, we are able to form models with sdB primaries at wide eccentric orbits.

\begin{table}
\caption{Parameter ranges used to determine the period-eccentricity distribution of the phase-dependent RLOF models and the CB-disk models. Not all parameter combinations result in an sdB binary.}\label{tb_parameter_ranges}
\centering
 \begin{tabular}{lr@{ - }l}
 \hline\hline
  Parameter	&	\multicolumn{2}{c}{Range} \\\hline
  \noalign{\smallskip}
  \multicolumn{3}{c}{Binary} \\
  sdB progenitor mass ($M_{\odot}$)		& 1.0 & 1.50 \\
  companion mass\tablefootmark{a} ($M_{\odot}$)	& 0.8 & 1.45 \\
  period (d) 					& 500 & 900 \\
  \multicolumn{3}{c}{Mass loss} \\
  $\alpha$			& 0.0 & 1.0 \\
  $\beta$			& 0.0 & 1.0 \\
  $\delta$			& 0.0 & 0.30 \\
  $\gamma$			& \multicolumn{2}{c}{1.12} \\
  $\epsilon$			& 0.0 & 0.25 \\
  \multicolumn{3}{c}{CB disk} \\
  maximum mass ($M_{\odot}$) 	& 10$^{-4}$ & 10$^{-2}$ 	\\
  lifetime	(yr)		& 10$^4$ & 10$^5$		\\
  viscosity		  	& 0.001 & 0.01			\\
  mass distribution		& \multicolumn{2}{c}{r$^{-1}$}	\\
  \hline
 \end{tabular}
 \tablefoot{\tablefoottext{a}{The companion mass is always lower as the donor mass.}}
\end{table}

From Fig. \ref{fig_period_ecc_model_coverage}, we can conclude that the region covered by the models with phase-dependent RLOF does not correspond with the observations. The effect of most parameters is to reduce the orbital period, while increasing the eccentricity, leading to an eccentricity pumping effect that is most efficient at short periods. At high orbital periods, the eccentricity pumping is too weak to overcome the circularisation, leading to circular orbits at periods over $\sim$1250 days. It is thus possible to recreate the short-period systems with moderate eccentricity, but the highly eccentric long-period systems cannot be reached. Another discrepancy is that the circular systems at orbital periods around 750 days cannot be reproduced either, when the eccentricity-pumping effects are active. The only way to model these circular systems at low orbital periods is by turning off all eccentricity-pumping mechanisms.

The models including a CB disk have a larger coverage than the models with only phase-dependent RLOF. However, even the disk models cannot completely reproduce the observed systems. They suffer the same problems as the models with only phase-dependent RLOF. Again, the effect of almost all parameters is to increase the eccentricity at the cost of lowering the orbital period. Even though higher periods and eccentricities can be reached, the system with the highest orbital period and eccentricity can still not be covered.

The period-eccentricity distribution of the two eccentricity pumping models seems to indicate that there is a trend of higher eccentricities at lower periods, opposite to the observed trend of higher eccentricities at higher periods. This is under the assumption that the method-depending parameters might vary independent of each other and the initial binary parameters. However, it is not unreasonable that for example the mass-loss fractions ($\alpha$, $\beta$ and $\delta$) depend on other binary properties like the mass ratio. Such a dependence between different parameters might indeed reproduce the observed period-eccentricity trend, but the models presented here can not prove nor disprove any such dependence.

\section{Summary and conclusions}\label{sec_conclusion}
The goal of this article was to test if three eccentricity-pumping mechanisms proposed in the literature, could recreate the observed period-eccentricity distribution of long-period binaries containing a hot subdwarf B star. The three proposed mechanisms are 1. tidally-enhanced wind mass-loss, 2. eccentricity pumping through phase-dependent RLOF and 3. the interaction between the binary and a circumbinary disk formed during RLOF combined with phase-dependent RLOF. The tidally-enhanced wind mass-loss method is not combined with any of the two other methods, as the point of the enhanced wind is to prevent the donor star from filling its Roche lobe, and in that way reducing the tidal forces. This would be counterproductive in the context of phase-dependent Roche-lobe overflow.

In this work we attempted to describe the effects of model parameters on the relationship between final period and eccentricity. Further work must be done in order to combine these models with realistic distributions of starting parameters in a binary-population-synthesis context.

Creating eccentric orbits with a tidally-enhanced-mass-loss mechanism is based on two processes. By increasing the wind mass loss, the radius of the donor star is kept well within its Roche lobe, thus reducing the tidal forces and maintaining the initial eccentricity of the orbit. The tidally-increased wind mass-loss depends on the orbital phase. This phase-dependent mass loss will further increase the eccentricity. The models calculated with MESA indeed lead to binaries on an eccentric orbit. However, the amount of mass that the donor star needs to lose is so large that the final mass of the donor star is lower than the minimum core mass necessary to ignite helium. All eccentric binaries formed through this channel end up as He-WDs, while all sdB stars ended on a circularised orbit as the wind-enhancement parameter needs to be very low. 

The phase-dependent RLOF mechanism is based on the same principle as the phase-dependent wind mass loss. By assuming a minimal eccentricity, the mass-loss rate will be higher during periastron than during apastron, and the eccentricity may increase. This method can create eccentric sdB binaries over a reasonable period-eccentricity range. The maximum eccentricity tends to be around 0.15, while the period range for eccentric sdBs is 600 to 1200 days. sdB binaries with longer orbital period, reaching $\sim$1600 days can be formed but are circularised within our parameter range. This mechanism can explain the observed systems with moderate eccentricities (0.05 $<$ $e$ $<$ 0.15) on shorter orbital periods ($P$ $<$ 1100 d). Neither the systems with higher eccentricities, nor the circularised low-period systems can be formed this way.

Adding CB disks to the models with phase-dependent RLOF does further increase the eccentricity of the produced binaries due to resonances between the binary and the CB disk. This binary-CB disk interaction increases the eccentricity while slightly decreasing the orbital period. These models can reach eccentricities up to 0.35 when assuming reasonable values for the disk parameters, and could potentially reach far higher eccentricities if more extreme values for maximum disk mass or surface mass distribution would be adopted. The disk model can reproduce most of the observed sdB binaries, except the circular short-period systems. 

When the models are compared to the observed sample, it is clear that phase-dependent RLOF in combination with CB disks can cover a significant part of the period-eccentricity diagram. However, there remain conflicts. The circular systems at short periods can not be formed, as well at the system with the higher period and eccentricity. Part of the discrepancy between the observed and modelled period-eccentricity diagram can be explained by insufficiently accurate models, while it is also possible that certain areas of the period-eccentricity diagram have not yet been populated due to a lack of observations. 

The observed trend of higher eccentricities at higher orbital periods does not follow directly from the method-depending parameters in the presented models. In our future research we will investigate whether certain parameters are dependent on the initial or current configuration of the binary. Such a dependence could potentially result in the observed period-eccentricity trend. However, the models presented here can neither prove nor disprove any such trend.

All three tested eccentricity-pumping mechanism are derived under certain assumptions, such as isotropic mass loss for eccentricity pumping through phase-dependent mass loss, or the approximation of a thin disk for the CB disk-binary interactions. Furthermore, values for several input parameters, for example the mass loss fractions ($\alpha$, $\beta$ and $\gamma$), are badly constrained or currently unknown, and are also likely to be functions of the system parameters such as mass ratio and companion mass rather than constants. While stable Keplerian discs are likely common in other evolved binary systems (see introduction), they are not yet documented around sdB wide binaries. We advocate here a specific search for circumstellar matter around wide sdB binaries.

In this article we have shown that we are able to form binary models with an sdB primary at wide eccentric orbits by eccentricity pumping of phase-dependent RLOF and CB disks. Tidally-enhanced wind mass-loss is unlikely to contribute to the formation of eccentric sdBs. Small discrepancies between the observed systems and the theoretical models remain, and will need to be addressed in future work.

\begin{acknowledgements}
  The research leading to these results has received funding from the European Research Council under the European Community's Seventh Framework Programme (FP7/2007--2013)/ERC grant agreement N$^{\underline{\mathrm o}}$\,227224 ({\sc prosperity}), as well as from the Research Council of K.U.Leuven grant agreements GOA/2008/04 and GOA/2013/012.
  The following Internet-based resources were used in research for this paper: the NASA Astrophysics Data System; the SIMBAD database and the VizieR service operated by CDS, Strasbourg, France; the ar$\chi$ive scientific paper preprint service operated by Cornell University. 
\end{acknowledgements}

\bibliographystyle{aa}
\bibliography{references}

\begin{thebibliography}{81}
\expandafter\ifx\csname natexlab\endcsname\relax\def\natexlab#1{#1}\fi

\bibitem[{{Artymowicz} \& {Lubow}(1994)}]{Artymowicz1994}
{Artymowicz}, P. \& {Lubow}, S.~H. 1994, \apj, 421, 651

\bibitem[{{Barlow} {et~al.}(2012){Barlow}, {Wade}, {Liss}, {{\O}stensen}, \&
  {Van Winckel}}]{Barlow12}
{Barlow}, B.~N., {Wade}, R.~A., {Liss}, S.~E., {{\O}stensen}, R.~H., \& {Van
  Winckel}, H. 2012, \apj, 758, 58

\bibitem[{{Blocker}(1995)}]{Blocker1995}
{Blocker}, T. 1995, \aap, 297, 727

\bibitem[{{Boffin} \& {Jorissen}(1988)}]{Boffin1988}
{Boffin}, H.~M.~J. \& {Jorissen}, A. 1988, \aap, 205, 155

\bibitem[{{Bona{\v c}i{\'c} Marinovi{\'c}} {et~al.}(2008){Bona{\v c}i{\'c}
  Marinovi{\'c}}, {Glebbeek}, \& {Pols}}]{Bonacic08}
{Bona{\v c}i{\'c} Marinovi{\'c}}, A.~A., {Glebbeek}, E., \& {Pols}, O.~R. 2008,
  \aap, 480, 797

\bibitem[{{Brassard} {et~al.}(2001){Brassard}, {Fontaine}, {Bill{\`e}res},
  {Charpinet}, {Liebert}, \& {Saffer}}]{Brassard01}
{Brassard}, P., {Fontaine}, G., {Bill{\`e}res}, M., {et~al.} 2001, \apj, 563,
  1013

\bibitem[{{Brown} {et~al.}(1997){Brown}, {Ferguson}, {Davidsen}, \&
  {Dorman}}]{Brown97}
{Brown}, T.~M., {Ferguson}, H.~C., {Davidsen}, A.~F., \& {Dorman}, B. 1997,
  \apj, 482, 685

\bibitem[{{Bujarrabal} {et~al.}(2005){Bujarrabal}, {Castro-Carrizo}, {Alcolea},
  \& {Neri}}]{Bujarrabal2005}
{Bujarrabal}, V., {Castro-Carrizo}, A., {Alcolea}, J., \& {Neri}, R. 2005,
  \aap, 441, 1031

\bibitem[{{Bujarrabal} {et~al.}(2015){Bujarrabal}, {Castro-Carrizo}, {Alcolea},
  \& {Van Winckel}}]{Bujarrabal2015}
{Bujarrabal}, V., {Castro-Carrizo}, A., {Alcolea}, J., \& {Van Winckel}, H.
  2015, ArXiv e-prints [\eprint[arXiv]{1502.01607}]

\bibitem[{{Bujarrabal} {et~al.}(2013){Bujarrabal}, {Castro-Carrizo}, {Alcolea},
  {Van Winckel}, {S{\'a}nchez Contreras}, {Santander-Garc{\'{\i}}a}, {Neri}, \&
  {Lucas}}]{Bujarrabal2013}
{Bujarrabal}, V., {Castro-Carrizo}, A., {Alcolea}, J., {et~al.} 2013, \aap,
  557, L11

\bibitem[{{Bujarrabal} {et~al.}(2007){Bujarrabal}, {van Winckel}, {Neri},
  {Alcolea}, {Castro-Carrizo}, \& {Deroo}}]{Bujarrabal2007}
{Bujarrabal}, V., {van Winckel}, H., {Neri}, R., {et~al.} 2007, \aap, 468, L45

\bibitem[{{Chen} {et~al.}(2013){Chen}, {Han}, {Deca}, \&
  {Podsiadlowski}}]{Chen13}
{Chen}, X., {Han}, Z., {Deca}, J., \& {Podsiadlowski}, P. 2013, \mnras, 434,
  186

\bibitem[{{Clausen} {et~al.}(2012){Clausen}, {Wade}, {Kopparapu}, \&
  {O'Shaughnessy}}]{Clausen2012}
{Clausen}, D., {Wade}, R.~A., {Kopparapu}, R.~K., \& {O'Shaughnessy}, R. 2012,
  \apj, 746, 186

\bibitem[{{Copperwheat} {et~al.}(2011){Copperwheat}, {Morales-Rueda}, {Marsh},
  {Maxted}, \& {Heber}}]{Copperwheat11}
{Copperwheat}, C.~M., {Morales-Rueda}, L., {Marsh}, T.~R., {Maxted}, P.~F.~L.,
  \& {Heber}, U. 2011, \mnras, 415, 1381

\bibitem[{{de Ruyter} {et~al.}(2005){de Ruyter}, {van Winckel}, {Dominik},
  {Waters}, \& {Dejonghe}}]{deRuyter2005}
{de Ruyter}, S., {van Winckel}, H., {Dominik}, C., {Waters}, L.~B.~F.~M., \&
  {Dejonghe}, H. 2005, \aap, 435, 161

\bibitem[{{de Ruyter} {et~al.}(2006){de Ruyter}, {van Winckel}, {Maas}, {Lloyd
  Evans}, {Waters}, \& {Dejonghe}}]{deRuyter2006}
{de Ruyter}, S., {van Winckel}, H., {Maas}, T., {et~al.} 2006, \aap, 448, 641

\bibitem[{{Deca} {et~al.}(2012){Deca}, {Marsh}, {{\O}stensen}, {Morales-Rueda},
  {Copperwheat}, {Wade}, {Stark}, {Maxted}, {Nelemans}, \& {Heber}}]{Deca12}
{Deca}, J., {Marsh}, T.~R., {{\O}stensen}, R.~H., {et~al.} 2012, \mnras, 421,
  2798

\bibitem[{{Deca} \& {Vos}(2015)}]{Deca2015}
{Deca}, J. \& {Vos}, J. 2015, to be published in \mnras

\bibitem[{{Dermine} {et~al.}(2013){Dermine}, {Izzard}, {Jorissen}, \& {Van
  Winckel}}]{Dermine2013}
{Dermine}, T., {Izzard}, R.~G., {Jorissen}, A., \& {Van Winckel}, H. 2013,
  \aap, 551, A50

\bibitem[{{Dullemond} \& {Monnier}(2010)}]{Dullemond2010}
{Dullemond}, C.~P. \& {Monnier}, J.~D. 2010, \araa, 48, 205

\bibitem[{{Eggleton}(2006)}]{Eggleton2006}
{Eggleton}, P. 2006, {Evolutionary Processes in Binary and Multiple Stars}

\bibitem[{{Eggleton}(1983)}]{Eggleton1983}
{Eggleton}, P.~P. 1983, \apj, 268, 368

\bibitem[{{Frank} {et~al.}(2002){Frank}, {King}, \& {Raine}}]{Frank2002}
{Frank}, J., {King}, A., \& {Raine}, D.~J. 2002, {Accretion Power in
  Astrophysics: Third Edition}

\bibitem[{{Geier} {et~al.}(2011){Geier}, {Hirsch}, {Tillich}, {Maxted},
  {Bentley}, {{\O}stensen}, {Heber}, {G{\"a}nsicke}, {Marsh}, {Napiwotzki},
  {Barlow}, \& {O'Toole}}]{Geier11}
{Geier}, S., {Hirsch}, H., {Tillich}, A., {et~al.} 2011, \aap, 530, A28

\bibitem[{{Gielen} {et~al.}(2011){Gielen}, {Bouwman}, {van Winckel}, {Lloyd
  Evans}, {Woods}, {Kemper}, {Marengo}, {Meixner}, {Sloan}, \&
  {Tielens}}]{Gielen2011}
{Gielen}, C., {Bouwman}, J., {van Winckel}, H., {et~al.} 2011, \aap, 533, A99

\bibitem[{{Gielen} {et~al.}(2008){Gielen}, {van Winckel}, {Min}, {Waters}, \&
  {Lloyd Evans}}]{Gielen2008}
{Gielen}, C., {van Winckel}, H., {Min}, M., {Waters}, L.~B.~F.~M., \& {Lloyd
  Evans}, T. 2008, \aap, 490, 725

\bibitem[{{Goldreich} \& {Tremaine}(1979)}]{Goldreich1979}
{Goldreich}, P. \& {Tremaine}, S. 1979, \apj, 233, 857

\bibitem[{{Green} {et~al.}(2001){Green}, {Liebert}, \& {Saffer}}]{Green01}
{Green}, E.~M., {Liebert}, J., \& {Saffer}, R.~A. 2001, in ASPCS, Vol. 226,
  12th European Workshop on White Dwarfs, ed. {J.~L.~Provencal, H.~L.~Shipman,
  J.~MacDonald, \& S.~Goodchild }, 192

\bibitem[{{Green} {et~al.}(1986){Green}, {Schmidt}, \& {Liebert}}]{Green86}
{Green}, R.~F., {Schmidt}, M., \& {Liebert}, J. 1986, \apjs, 61, 305

\bibitem[{{Greggio} \& {Renzini}(1990)}]{Greggio90}
{Greggio}, L. \& {Renzini}, A. 1990, \apj, 364, 35

\bibitem[{{Han} {et~al.}(2003){Han}, {Podsiadlowski}, {Maxted}, \&
  {Marsh}}]{Han03}
{Han}, Z., {Podsiadlowski}, P., {Maxted}, P.~F.~L., \& {Marsh}, T.~R. 2003,
  \mnras, 341, 669

\bibitem[{{Han} {et~al.}(2002){Han}, {Podsiadlowski}, {Maxted}, {Marsh}, \&
  {Ivanova}}]{Han02}
{Han}, Z., {Podsiadlowski}, P., {Maxted}, P.~F.~L., {Marsh}, T.~R., \&
  {Ivanova}, N. 2002, \mnras, 336, 449

\bibitem[{{Han} {et~al.}(2000){Han}, {Tout}, \& {Eggleton}}]{Han00}
{Han}, Z., {Tout}, C.~A., \& {Eggleton}, P.~P. 2000, \mnras, 319, 215

\bibitem[{{Heber}(1998)}]{Heber98}
{Heber}, U. 1998, in ESA Special Publication, Vol. 413, Ultraviolet
  Astrophysics Beyond the IUE Final Archive, ed. W.~{Wamsteker}, R.~{Gonzalez
  Riestra}, \& B.~{Harris}, 195

\bibitem[{{Heber}(2009)}]{Heber09}
{Heber}, U. 2009, \araa, 47, 211

\bibitem[{{Heber} {et~al.}(2002){Heber}, {Moehler}, {Napiwotzki}, {Thejll}, \&
  {Green}}]{Heber02}
{Heber}, U., {Moehler}, S., {Napiwotzki}, R., {Thejll}, P., \& {Green}, E.~M.
  2002, \aap, 383, 938

\bibitem[{{Hillen}(2015)}]{Hillen2015}
{Hillen}, M. 2015, to be published in \aap

\bibitem[{{Hillen} {et~al.}(2014){Hillen}, {Menu}, {Van Winckel}, {Min},
  {Gielen}, {Wevers}, {Mulders}, {Regibo}, \& {Verhoelst}}]{Hillen2014}
{Hillen}, M., {Menu}, J., {Van Winckel}, H., {et~al.} 2014, \aap, 568, A12

\bibitem[{{Hurley} {et~al.}(2002){Hurley}, {Tout}, \& {Pols}}]{Hurley2002}
{Hurley}, J.~R., {Tout}, C.~A., \& {Pols}, O.~R. 2002, \mnras, 329, 897

\bibitem[{{Hut}(1981)}]{Hut1981}
{Hut}, P. 1981, \aap, 99, 126

\bibitem[{{Kamath} {et~al.}(2014){Kamath}, {Wood}, \& {Van
  Winckel}}]{Kamath2014}
{Kamath}, D., {Wood}, P.~R., \& {Van Winckel}, H. 2014, \mnras, 439, 2211

\bibitem[{{Kamath} {et~al.}(2015){Kamath}, {Wood}, \& {Van
  Winckel}}]{Kamath2015}
{Kamath}, D., {Wood}, P.~R., \& {Van Winckel}, H. 2015, to be published in
  \mnras

\bibitem[{{Koen} {et~al.}(1998){Koen}, {Orosz}, \& {Wade}}]{Koen98}
{Koen}, C., {Orosz}, J.~A., \& {Wade}, R.~A. 1998, \mnras, 300, 695

\bibitem[{{Kolb} \& {Ritter}(1990)}]{Kolb1990}
{Kolb}, U. \& {Ritter}, H. 1990, \aap, 236, 385

\bibitem[{{Ku{\v c}inskas}(1998)}]{Kucinskas1998}
{Ku{\v c}inskas}, A. 1998, \apss, 262, 127

\bibitem[{{Lubow} \& {Artymowicz}(1996)}]{Lubow1996}
{Lubow}, S.~H. \& {Artymowicz}, P. 1996, in NATO Advanced Science Institutes
  (ASI) Series C, Vol. 477, NATO Advanced Science Institutes (ASI) Series C,
  ed. R.~A.~M.~J. {Wijers}, M.~B. {Davies}, \& C.~A. {Tout}, 53

\bibitem[{{Lubow} \& {Artymowicz}(2000)}]{Lubow2000}
{Lubow}, S.~H. \& {Artymowicz}, P. 2000, Protostars and Planets IV, 731

\bibitem[{{Maxted} {et~al.}(2001){Maxted}, {Heber}, {Marsh}, \&
  {North}}]{Maxted01}
{Maxted}, P.~f.~L., {Heber}, U., {Marsh}, T.~R., \& {North}, R.~C. 2001,
  \mnras, 326, 1391

\bibitem[{{Maxted} {et~al.}(2000){Maxted}, {Moran}, {Marsh}, \&
  {Gatti}}]{Maxted00}
{Maxted}, P.~F.~L., {Moran}, C.~K.~J., {Marsh}, T.~R., \& {Gatti}, A.~A. 2000,
  \mnras, 311, 877

\bibitem[{{Meyer} \& {Meyer-Hofmeister}(1983)}]{Meyer1983}
{Meyer}, F. \& {Meyer-Hofmeister}, E. 1983, \aap, 121, 29

\bibitem[{{Morales-Rueda} {et~al.}(2003){Morales-Rueda}, {Maxted}, {Marsh},
  {North}, \& {Heber}}]{Morales03}
{Morales-Rueda}, L., {Maxted}, P.~F.~L., {Marsh}, T.~R., {North}, R.~C., \&
  {Heber}, U. 2003, \mnras, 338, 752

\bibitem[{{Napiwotzki} {et~al.}(2004){Napiwotzki}, {Karl}, {Lisker}, {Heber},
  {Christlieb}, {Reimers}, {Nelemans}, \& {Homeier}}]{Napiwotzki04}
{Napiwotzki}, R., {Karl}, C.~A., {Lisker}, T., {et~al.} 2004, \apss, 291, 321

\bibitem[{{Nelemans} {et~al.}(2000){Nelemans}, {Verbunt}, {Yungelson}, \&
  {Portegies Zwart}}]{Nelemans2000}
{Nelemans}, G., {Verbunt}, F., {Yungelson}, L.~R., \& {Portegies Zwart}, S.~F.
  2000, \aap, 360, 1011

\bibitem[{{{\O}stensen}(2014)}]{Oestensen2014}
{{\O}stensen}, R.~H. 2014, in Astronomical Society of the Pacific Conference
  Series, Vol. 481, 6th Meeting on Hot Subdwarf Stars and Related Objects, ed.
  V.~{van Grootel}, E.~{Green}, G.~{Fontaine}, \& S.~{Charpinet}, 37

\bibitem[{{{\O}stensen} {et~al.}(2012){{\O}stensen}, {Degroote}, {Telting},
  {Vos}, {Aerts}, {Jeffery}, {Green}, {Reed}, \& {Heber}}]{Oestensen2012b}
{{\O}stensen}, R.~H., {Degroote}, P., {Telting}, J.~H., {et~al.} 2012, \apjl,
  753, L17

\bibitem[{{{\O}stensen} \& {Van Winckel}(2011)}]{Oestensen2011}
{{\O}stensen}, R.~H. \& {Van Winckel}, H. 2011, in ASPCS, Vol. 447, Evolution
  of Compact Binaries, ed. {L.~Schmidtobreick, M.~R.~Schreiber, \& C.~Tappert},
  171

\bibitem[{{{\O}stensen} \& {Van Winckel}(2012)}]{Oestensen2012}
{{\O}stensen}, R.~H. \& {Van Winckel}, H. 2012, in ASPCS, Vol. 452, Fifth
  Meeting on Hot Subdwarf Stars and Related Objects, ed. {D.~Kilkenny,
  C.~S.~Jeffery, \& C.~Koen}, 163

\bibitem[{{Paczynski}(1976)}]{Paczynski76}
{Paczynski}, B. 1976, in IAU Symposium, Vol.~73, Structure and Evolution of
  Close Binary Systems, ed. P.~{Eggleton}, S.~{Mitton}, \& J.~{Whelan}, 75

\bibitem[{{Paxton} {et~al.}(2011){Paxton}, {Bildsten}, {Dotter}, {Herwig},
  {Lesaffre}, \& {Timmes}}]{Paxton2011}
{Paxton}, B., {Bildsten}, L., {Dotter}, A., {et~al.} 2011, \apjs, 192, 3

\bibitem[{{Paxton} {et~al.}(2013){Paxton}, {Cantiello}, {Arras}, {Bildsten},
  {Brown}, {Dotter}, {Mankovich}, {Montgomery}, {Stello}, {Timmes}, \&
  {Townsend}}]{Paxton2013}
{Paxton}, B., {Cantiello}, M., {Arras}, P., {et~al.} 2013, \apjs, 208, 4

\bibitem[{{Pennington}(1985)}]{Pennington1985}
{Pennington}, R. 1985, {Interacting binary stars: appendix.}, ed. J.~E.
  {Pringle} \& R.~A. {Wade}, 197--199

\bibitem[{{Rasio} {et~al.}(1996){Rasio}, {Tout}, {Lubow}, \&
  {Livio}}]{Rasio1996}
{Rasio}, F.~A., {Tout}, C.~A., {Lubow}, S.~H., \& {Livio}, M. 1996, \apj, 470,
  1187

\bibitem[{{Raskin} {et~al.}(2011){Raskin}, {van Winckel}, {Hensberge},
  {Jorissen}, {Lehmann}, {Waelkens}, {Avila}, {de Cuyper}, {Degroote},
  {Dubosson}, {Dumortier}, {Fr{\'e}mat}, {Laux}, {Michaud}, {Morren}, {Perez
  Padilla}, {Pessemier}, {Prins}, {Smolders}, {van Eck}, \&
  {Winkler}}]{Raskin2011}
{Raskin}, G., {van Winckel}, H., {Hensberge}, H., {et~al.} 2011, \aap, 526, A69

\bibitem[{{Reimers}(1975)}]{Reimers1975}
{Reimers}, D. 1975, {Circumstellar envelopes and mass loss of red giant stars},
  ed. B.~{Baschek}, W.~H. {Kegel}, \& G.~{Traving}, 229--256

\bibitem[{{Ritter}(1988)}]{Ritter1988}
{Ritter}, H. 1988, A\&A, 202, 93

\bibitem[{{Roedig} {et~al.}(2011){Roedig}, {Dotti}, {Sesana}, {Cuadra}, \&
  {Colpi}}]{Roedig2011}
{Roedig}, C., {Dotti}, M., {Sesana}, A., {Cuadra}, J., \& {Colpi}, M. 2011,
  \mnras, 415, 3033

\bibitem[{{Saffer} {et~al.}(1994){Saffer}, {Bergeron}, {Koester}, \&
  {Liebert}}]{Saffer94}
{Saffer}, R.~A., {Bergeron}, P., {Koester}, D., \& {Liebert}, J. 1994, \apj,
  432, 351

\bibitem[{{Siess} {et~al.}(2014){Siess}, {Davis}, \& {Jorissen}}]{Siess2014}
{Siess}, L., {Davis}, P.~J., \& {Jorissen}, A. 2014, \aap, 565, A57

\bibitem[{{Soberman} {et~al.}(1997){Soberman}, {Phinney}, \& {van den
  Heuvel}}]{Soberman1997}
{Soberman}, G.~E., {Phinney}, E.~S., \& {van den Heuvel}, E.~P.~J. 1997, \aap,
  327, 620

\bibitem[{{Soker}(2000)}]{Soker2000}
{Soker}, N. 2000, \aap, 357, 557

\bibitem[{{Soker}(2014)}]{Soker2014}
{Soker}, N. 2014, ArXiv e-prints [\eprint[arXiv]{1410.5363}]

\bibitem[{{Tauris} \& {van den Heuvel}(2006)}]{Tauris2006}
{Tauris}, T.~M. \& {van den Heuvel}, E.~P.~J. 2006, {Formation and evolution of
  compact stellar X-ray sources}, ed. W.~H.~G. {Lewin} \& M.~{van der Klis},
  623--665

\bibitem[{{Tout} \& {Eggleton}(1988)}]{Tout1988}
{Tout}, C.~A. \& {Eggleton}, P.~P. 1988, \mnras, 231, 823

\bibitem[{{van Aarle} {et~al.}(2011){van Aarle}, {van Winckel}, {Lloyd Evans},
  {Ueta}, {Wood}, \& {Ginsburg}}]{vanAarle2011}
{van Aarle}, E., {van Winckel}, H., {Lloyd Evans}, T., {et~al.} 2011, \aap,
  530, A90

\bibitem[{{Vos} {et~al.}(2014){Vos}, {{\"O}stensen}, \& {Van Winckel}}]{Vos14}
{Vos}, J., {{\"O}stensen}, R., \& {Van Winckel}, H. 2014, in Astronomical
  Society of the Pacific Conference Series, Vol. 481, 6th Meeting on Hot
  Subdwarf Stars and Related Objects, ed. V.~{van Grootel}, E.~{Green},
  G.~{Fontaine}, \& S.~{Charpinet}, 265

\bibitem[{{Vos} {et~al.}(2012){Vos}, {{\O}stensen}, {Degroote}, {De Smedt},
  {Green}, {Heber}, {Van Winckel}, {Acke}, {Bloemen}, {De Cat}, {Exter},
  {Lampens}, {Lombaert}, {Masseron}, {Menu}, {Neyskens}, {Raskin}, {Ringat},
  {Rauch}, {Smolders}, \& {Tkachenko}}]{Vos12}
{Vos}, J., {{\O}stensen}, R.~H., {Degroote}, P., {et~al.} 2012, \aap, 548, A6

\bibitem[{{Vos} {et~al.}(2013){Vos}, {{\O}stensen}, {N{\'e}meth}, {Green},
  {Heber}, \& {Van Winckel}}]{Vos13}
{Vos}, J., {{\O}stensen}, R.~H., {N{\'e}meth}, P., {et~al.} 2013, \aap, 559,
  A54

\bibitem[{{Webbink}(1984)}]{Webbink84}
{Webbink}, R.~F. 1984, \apj, 277, 355

\bibitem[{{Zahn}(1975)}]{Zahn1975}
{Zahn}, J.-P. 1975, \aap, 41, 329

\bibitem[{{Zahn}(1977)}]{Zahn1977}
{Zahn}, J.-P. 1977, \aap, 57, 383

\bibitem[{{Zahn}(2005)}]{Zahn2005}
{Zahn}, J.-P. 2005, in Astronomical Society of the Pacific Conference Series,
  Vol. 333, Tidal Evolution and Oscillations in Binary Stars, ed. A.~{Claret},
  A.~{Gim{\'e}nez}, \& J.-P. {Zahn}, 4

\end{thebibliography}

\begin{appendix}
 
\section{Binary physics in MESA}\label{app_mesa}
In this appendix we will list the physics used in the binary evolution module of MESA. For the exact software implementation we refer the reader to the open-source code, and future instrument papers.

\subsection{Evolution of the orbital parameters}
The orbital angular momentum of a binary system is:
\begin{equation}
 J_{\mathrm{orb}} = M_\dsub \cdot M_\asub \sqrt{ \frac{G  a (1-e^2)}{M_\dsub + M_\asub} }. \label{eq_jorb-sep}
\end{equation}
Where $a$ is the binary separation, and $e$ is the eccentricity of the orbit. $M_d$ and $M_a$ are respectively the mass of the donor and the accretor, and $G$ is the gravitational constant. The evolution of the binary separation is obtained from the time derivative of Eq. (\ref{eq_jorb-sep}):
\begin{equation}
 \frac{\dot{a}}{a} = 2 \frac{\dot{J}_{\mathrm{orb}}}{J_{\mathrm{orb}}} - 2 \frac{\dot{M}_\dsub}{M_\dsub} - 2 \frac{\dot{M}_\asub}{M_\asub} + \frac{\dot{M}_\dsub + \dot{M}_\asub}{M_\dsub + M_\asub} + 2 \frac{e \dot{e}}{1-e^2}.
\end{equation}
In MESA, the mass loss and accretion rates, the change in orbital angular momentum, and the change in eccentricity are computed, and then used to update the orbital separation.

The change in mass of both components depends on the mass lost in stellar winds, the mass lost through Roche-lobe overflow (RLOF), and the fraction of mass that is accreted by the companion. For the donor star the net mass loss is:
\begin{equation}
 \dot{M}_\dsub = \dot{M}_{\mathrm{rlof}} + \dot{M}_{\mathrm{wind,d}} - \epsilon_{\mathrm{wind,a}} \cdot \dot{M}_{\mathrm{wind,a}}, \label{eq_mdot_donor}
\end{equation}
while for the accretor:
\begin{equation}
 \dot{M}_\asub = \dot{M}_{\mathrm{wind,a}} - \epsilon_{\mathrm{wind,d}} \cdot \dot{M}_{\mathrm{wind,d}} - \epsilon_{\mathrm{rlof}} \cdot \dot{M}_{\mathrm{rlof}}.
\end{equation}
Where $\epsilon_{\mathrm{wind}}$ is the fraction of the wind mass loss that is accreted by the companion and $\epsilon_{\mathrm{rlof}}$ is the fraction of the RLOF mass loss accreted by the companion. The amount of mass lost to infinity is then:
\begin{multline}
\dot{M}_{\infty} = 
    (1 - \epsilon_{\mathrm{rlof}}) \cdot \dot{M}_{\mathrm{rlof}} 
    + (1 - \epsilon_{\mathrm{wind,d}}) \cdot \dot{M}_{\mathrm{wind,d}} \\
    + (1 - \epsilon_{\mathrm{wind,a}}) \cdot \dot{M}_{\mathrm{wind,a}}.
\end{multline}

The change in orbital angular momentum has three main components: mass lost to infinity in stellar winds, mass lost to infinity during Roche-lobe overflow and the change in angular momentum due to resonances between the binary and a potential circumbinary (CB) disk:
\begin{equation}
 \dot{J}_{\mathrm{orb}} = \dot{J}_{\mathrm{rlof}} + \dot{J}_{\mathrm{wind}} + \dot{J}_{\mathrm{disk}}.
\end{equation}
In the wide binary models considered here, the strongest contributions to $\dot{J}_{\mathrm{orb}}$ will come from mass loss during Roche-lobe overflow and the interaction between the binary and a CB disk. 

The evolution of the eccentricity is governed by the tidal interactions between both stars, the effect of phase-dependent mass loss/transfer which can pump or decrease the eccentricity based on the mass ratio and the interaction between the binary and a circumbinary disk:
\begin{equation}
 \dot{e} = \dot{e}_{\mathrm{tides}} + \dot{e}_{\mathrm{ml}} + \dot{e}_{\mathrm{disk}}. \label{eq_edot}
\end{equation}

The relation between the orbital period and the separation is:
\begin{equation}
 P = 2 \pi \sqrt{ \frac{a^3}{G (M_\dsub + M_\asub)} }.
\end{equation}

All physical mechanisms relevant to the terms given in equations (\ref{eq_mdot_donor} - \ref{eq_edot}) are explained in the following subsections.

\subsection{Roche-lobe overflow}\label{app_rlof_mesa}
The instantaneous Roche-lobe radius of the donor star ($R_{\mathrm{L,d}}$) at a specific moment in the orbit is approximated by adapting the fitting formula for the effective Roche-lobe radius of \citet{Eggleton1983} by allowing the binary separation to  depend on the orbital phase ($\theta$):
\begin{equation}
 R_{\mathrm{L,d}}(\theta) = \frac{1 - e^2}{1 + e \cos(\theta)} \cdot \frac{ 0.49\,q_\dsub^{2/3}\ a }{ 0.60\,q_\dsub^{2/3} + \ln(1 + q_\dsub^{1/3}) } \label{eq_rochelobe}.
\end{equation}
Where $a$ is the separation, $e$ the orbital eccentricity. The mass ratios $q_\dsub$ and $q_\asub$ are defined as:
\begin{equation}
 q_{\dsub} = \frac{M_{\dsub}}{M_{\asub}},\ \ q_{\asub} = \frac{M_{\asub}}{M_{\dsub}}.
\end{equation}
The Roche-lobe radius of the accretor can be obtained by inverting the mass ratio in Eq. \ref{eq_rochelobe}, thus replacing $q_\dsub$ with $q_\asub$.

Mass-loss is implemented in MESA following the prescription of \citet{Ritter1988} and \citet{Kolb1990}. This formalism differs between mass-loss from the optically thin region ($R_\dsub \le R_{\mathrm{L,d}}$), and that from the optically thick region ($R_\dsub > R_{\mathrm{L,d}}$). Mass-loss from the former region is given by:
\begin{equation}
 \dot{M}_{\mathrm{thin}} = \dot{M}_0 \cdot \exp \left( \frac{R_\dsub - R_{\mathrm{L,d}}}{H_{P,{\rm L1}}} \right).
\end{equation}
Where $R_\dsub$ is the donor radius and $H_{P,{\rm L1}}$ is the pressure scale height at the inner Lagrange point (L1), which can be linked to the pressure scale height at the photosphere of the donor $H_{P,\mathrm{ph}}$ as:
\begin{equation}
 H_{P,{\rm L1}} = \frac{H_{P,\mathrm{ph}}}{\gamma(q)} = \frac{1}{\gamma(q)} \cdot \frac{P_{\mathrm{ph}}\, R_\dsub^2}{\rho_{\mathrm{ph}}\,G\,M_\dsub}.
\end{equation}
Here $P_{\mathrm{ph}}$ is the pressure at the donor photosphere and $\rho_{\mathrm{ph}}$ is the photospheric density of the donor. $\gamma(q)$ is a factor that depends on the mass ratio as:
\begin{equation}
 \gamma(q) = 
  \begin{cases}
    0.954 + 0.025\,\log_{10}(q) - 0.038\,\log_{10}^2(q), & 0.04 < q < 1.\\
    \\
    0.954 + 0.039\,\log_{10}(q) + 0.114\,\log_{10}^2(q), & 1 \le q \le 20.
  \end{cases}
\end{equation}
$\dot{M}_0$ is the mass loss rate when the donor star just fills its Roche-lobe:
\begin{equation}
 \dot{M}_0 = \frac{2 \pi}{\sqrt{e}} \left( \frac{\mathcal{R} T_{\mathrm{eff,d}}}{\mu_{\mathrm{ph,d}}} \right)^{3/2} \frac{R_{\mathrm{L,d}}^3}{G M_\dsub} \rho_{\mathrm{ph,\dsub}}\,F(q_\dsub). \label{eq_mdot_zero}
\end{equation}
$\mu_{\mathrm{ph,d}}$ and $\rho_{\mathrm{ph,d}}$ are the mean molecular weight and density at the donor photosphere, $T_{\mathrm{eff,d}}$ is the effective temperature of the donor and  $\mathcal{R}$ is the ideal gas constant. $F(q)$ is a function of the mass ratio depending on the Roche geometry, given by \citep[see also][]{Meyer1983}:
\begin{equation}
 F(q) = \frac{q}{\sqrt{ g^2(q) - (1+q)\,g(q)}}\,\left(\frac{a}{R_{\mathrm{L,d}}}\right)^3,
\end{equation}
where $g(q)$ is:
\begin{equation}
 g(q) = \frac{q}{x_{\mathrm{L}}^3} + \frac{1}{(1 - x_{\mathrm{L}})^3}.
\end{equation}
An approximation of the distance between the centre of mass of the donor and the L1 point in units of the binary separation ($x_L$) is given by \citet{Frank2002}:
\begin{equation}
 x_{\rm L} = 0.5 - 0.227 \log_{10}(q).
\end{equation}

In the case that the stellar radius significantly overfills its Roche-lobe, the Roche-lobe will lie in the optically thick region of the donor. The mass loss is then calculated via:
\begin{equation}
\dot{M}_{\mathrm{thick}} = \dot{M}_0 + 2 \pi F(q) \frac{R_{\mathrm{L,d}}^3}{G M_\dsub} \int_{P_{\rm L1}}^{P_{ph}} F_3(\Gamma_1) \sqrt{\frac{\mathcal{R}T}{\mu}} {\rm d}P.
\end{equation}
Where $\dot{M}_0$ is given in Eq. (\ref{eq_mdot_zero}), $P_{\rm L1}$ and $P_{ph}$ are respectively the pressure at L1 and at the stellar photosphere. Both the temperature $T$ and the mean molecular weight $\mu$ depend on the pressure. $F_3$ is a function that depends on the adiabatic exponent $\Gamma_1$, and is given by:
\begin{equation}
 F_3(\Gamma_1) = \sqrt{ \Gamma_1 } \left( \frac{2}{\Gamma_1 + 1} \right)^{\dfrac{\Gamma_1 + 1}{2 \Gamma_1 - 2}}.
\end{equation}

The mass loss equations depend on the eccentric anomaly through the Roche-lobe radius. The time steps that MESA uses in the evolution are much larger than one orbital period, thus to obtain the average mass loss rate to be used in a time step, the above equations are integrated over the orbit.

RLOF is not necessarily conservative. To allow for mass to be lost from the system the formalism of \citet{Tauris2006} and \citet{Soberman1997} is used. This system describes three fractions ($\alpha, \beta, \delta$) to lose mass from the system:
\begin{itemize}
 \item $\alpha$: mass lost from the vicinity of the donor as a fast wind (Jeans mode). This is modelled as a spherically symmetric outflow from the donor star in the form of a fast wind. The mass lost in this way carries the specific angular momentum of the donor star.
 \item $\beta$: mass lost from the vicinity of the accretor as a fast wind (Isotropic re-emission). A flow in which matter is transported from the donor to the vicinity of the accretor, where it is ejected as a fast isotropic wind. Mass lost in this way carries the specific angular momentum of the accretor.
 \item $\delta$: mass lost from a circumbinary coplanar toroid. The radius of the coplanar toroid is determined by $\gamma$ as $R_{\mathrm{toroid}} = \gamma^2 a$.
\end{itemize}
The accretion efficiency of RLOF is then given by:
\begin{equation}
 \epsilon_{\mathrm{rlof}} = 1 - \alpha - \beta - \delta.
\end{equation}

When the mode of mass loss is known, the change in angular momentum can be calculated. Mass accreted by the companion will not change the total angular momentum of the system, thus only fractions $\alpha, \beta$ and $\delta$ will have an influence on $\dot{J}$. The total effect of mass loss through Roche-lobe overflow on the change in angular momentum is given by \citet{Tauris2006}:
\begin{equation}
 \dot{J}_{\mathrm{rlof}} = \frac{\alpha + \beta q^2 + \delta \gamma (1+q)^2}{1 + q} \frac{\dot{M}_{\mathrm{rlof}}}{M_\dsub} \cdot J_{\mathrm{orb}}. \label{eq_mdot_jdot_tauris}
\end{equation}

\subsection{Wind mass loss}\label{app_wind_ml_mesa}
The amount of mass lost through stellar winds is determined by any of the wind loss prescriptions defined in the stellar part of MESA. The binary module offers the possibility to boost this mass loss using the Companion Reinforced Attrition Process (CRAP) mechanism \citep{Tout1988}, also known as tidally enhanced mass loss. In this model it is assumed that tidal interactions or magnetic activity are responsible for the enhancement of the wind. The enhancement is expected to have a similar dependence on radius over Roche-lobe radius as the torque in a tidal friction model, and there is an expected saturation when co-rotation is reached (in their model when $R = R_{\rm L}/2$).
\begin{equation}\label{eq_tout_crap}
 \dot{M}_{\mathrm{wind}} = \dot{M}_{\mathrm{Reimers}} \cdot \left\{ 1 + B_{\mathrm{wind}} \cdot \min\left[ \left(\frac{R}{R_{\rm L}}\right)^6, \frac{1}{2^6} \right] \right\}
\end{equation}
 The factor $B_{\mathrm{wind}}$ is estimated by \citet{Tout1988} to be of the order $B_{\mathrm{wind}} \approx 10^4$, but can vary significantly depending on which system needs to be explained.

Part of the mass lost due to stellar winds is accreted by the companion. In the case of fast winds ($v_{\mathrm{wind}} \gg v_{\mathrm{orb}}$) the accretion fraction is given by the Bondi-Hoyle mechanism \citep{Hurley2002}. Here we give the equations for the donor, those of the accretor can be obtained by switching the d and a subscripts:
\begin{equation} \label{eq_BH_accretion_fraction}
 \epsilon_{\mathrm{BH,d}} = \frac{1}{\sqrt{1 - e^2}} \left( \frac{G M_\dsub}{v_{\mathrm{wind,a}}^2} \right)^2 \frac{\alpha_{\rm BH}}{2 a^2} \frac{1}{(1+ v^2)^{3/2}}.
\end{equation}
The velocities are:
\begin{align}
 v^2 &= \frac{v_{\mathrm{orb}}^2}{v_{\mathrm{wind,a}}^2},\\
 v_{\mathrm{orb}}^2 &= \frac{G (M_\dsub + M_\asub)}{a},\\
 v_{\mathrm{wind,a}}^2 &= 2\,\beta_W\,\frac{G M_\asub}{R_\asub}.
\end{align}
The wind velocity used here is set proportional to the escape velocity from the stellar surface. Based on observed wind velocities in cool supergiants ($v_{\mathrm{wind}}=5 - 35$\,km\,s$^{-1}$) \citep{Kucinskas1998}, $\beta_W$ is taken to be 1/8. The free parameter $\alpha_{\rm BH}$ is set to 3/2 based on \citet{Boffin1988}.

The angular-momentum loss due to wind-mass loss, assuming a spherical-symmetric wind is:
\begin{equation}
 \dot{J}_{\mathrm{wind,d}} = \dot{M}_{\mathrm{wind,d}} \left( \frac{M_\asub}{M_\dsub + M_\asub} a \right)^2 \frac{2 \pi}{P} \sqrt{1 - e^2}. \label{eq_jdot_wind}
\end{equation}
For the change in total angular momentum of the binary due to stellar winds, the contribution of the accretor need to be added and $\dot{J}_{\mathrm{wind}} = \dot{J}_{\mathrm{wind,d}} + \dot{J}_{\mathrm{wind,a}}$.

\subsection{phase-dependent mass loss}\label{app_phase_dependent_ml_mesa}
When mass loss, weather due to stellar winds, RLOF or other reasons, is not constant during the binary orbit, it will have an effect on the eccentricity of the system. There are multiple causes of the periodicity of mass loss and accretion. For example, the mass loss might be caused by stellar pulsations, which can transfer more mass at the periastra that coincide with a maximum stellar radius. The methods implemented in MESA are phase-dependent wind-mass loss through tidal interactions, and phase-dependent RLOF on eccentric orbits. The latter will boost mass loss near the periastron passage while during apastron the star is completely inside its Roche lobe. The effect of phase-dependent mass loss on the orbital eccentricity was studied by \citet{Soker2000} based on the theoretical work of \citet[Ch. 6.5]{Eggleton2006}. 

In calculating the effect on the eccentricity we have to distinguish between mass lost to infinity, and mass that is accreted by the companion. Assuming isotropic mass loss, the change in eccentricity for mass lost from the system is given by:
\begin{equation}
\dot{e}_{\mathrm{lost}}(\theta) = \frac{|\dot{M}_{\infty}(\theta)|}{M_\dsub + M_\asub} ( e + \cos{\theta} ). \label{eq_edot_ml_inf}
\end{equation}
Where $\theta$ is the true anomaly, and $\dot{M}_{\infty}(\theta)$ is the mass lost from the system at a specific phase $\theta$ during the orbit. From this equation one can see that constant mass loss will not change the eccentricity, if however the fraction of mass lost near periastron is significantly larger than during the remainder of the orbit, Eq. \ref{eq_edot_ml_inf} predicts a positive $\dot{e}$.

The effect on $\dot{e}$ of mass accreted by the companion, again under the assumption of isotropic mass-loss, is given by:
\begin{equation}
 \dot{e}_{\mathrm{acc}}(\theta) = 2 |\dot{M}_{\mathrm{acc}}(\theta)| \left( \frac{1}{M_\dsub} - \frac{1}{M_\asub} \right) ( e + \cos{\theta} ). \label{eq_edot_ml_trans}
\end{equation}
Where $\dot{M}_{\mathrm{acc}}(\theta)$ is the mass accreted by the companion at a specific phase $\theta$ during the orbit. As mass transfer is expected near periastron, the eccentricity will increase if $M_\dsub$ < $M_\asub$, which is required in the case of stable mass transfer.

To obtain the total change in eccentricity in one orbit, Eq. \ref{eq_edot_ml_inf} and \ref{eq_edot_ml_trans} are integrated over the orbit:
\begin{equation}
 \dot{e}{_\mathrm{ml}} = \int_\theta [\,\dot{e}_{\mathrm{lost}}(\theta) + \dot{e}_{\mathrm{acc}}(\theta)\,]\ {\rm d}\theta.
\end{equation}

\subsection{Tidal forces}
The gravitational forces acting on the components in binary systems induce a deformation of their structure, and creates tidal bulges. These bulges are lagging behind, thus the gravitational attraction generates a torque on those bulges, which forces the synchronisation and circularisation of the stellar and orbital rotation. The orbital parameters change because the star's rotational energy is dissipated into heat. There are two cases one has to consider, stars with convective envelopes and stars with radiative envelopes. In the former, the kinetic energy of tidally-induced large-scale currents is dissipated into heat by viscous friction of the convective environment. In radiative stars it is mainly radiative damping of gravity modes that functions as dissipation process. See \citet{Zahn2005} for a review on tidal dissipation.

We use the formalism developed by \citet{Hut1981} to calculate the effect of tides on circularisation and synchronisation. This formalism, developed based on the weak-friction model, results in (where we again give only the equations relevant for the donor star, and those of the accretor are obtained by switching the d and a subscripts):
\begin{multline}
 \dot{e}_{\mathrm{tides,d}} = -27 \left(\frac{k}{T}\right)_\dsub q_\dsub(1+q_\dsub) \left(\frac{R_\dsub}{a}\right)^8 \frac{e}{(1-e^2)^{13/2}} \\
  \left( f_3(e^2) - \frac{11}{18}(1-e^2)^{3/2}f_4(e^2)\frac{\Omega}{\omega} \right)\ \ {\rm yr}^{-1},
\end{multline}
for the circularisation, while the synchronisation is given by:
\begin{multline}
 \dot{\Omega}_{\mathrm{tides,d}} = 3 \left(\frac{k}{T}\right)_\dsub \frac{q_\dsub^2}{r_{\rm g}^2}\left(\frac{R_\dsub}{a}\right)^8 \frac{\sqrt{G (M_\dsub M_\asub) / a^3}}{(1-e^2)^6} \\
  \left( f_2(e^2) - (1-e^2)^{3/2}f_5(e^2)\frac{\Omega}{\sqrt{G (M_\dsub M_\asub) / a^3}} \right)\ \ {\rm yr}^{-1}.
\end{multline}
Here $f_{2-5}$ are polynomials in e:
\begin{align}
 f_2(e^2) &= 1 + \frac{15}{2}e^2 + \frac{45}{8}e^4 + \frac{5}{16}e^6 , \\
 f_3(e^2) &= 1 + \frac{15}{4}e^2 + \frac{15}{8}e^4 + \frac{5}{64}e^6 , \\
 f_4(e^2) &= 1 + \frac{3}{2}e^2 + \frac{1}{8}e^4 , \\
 f_5(e^2) &= 1 + 3\,e^2 + \frac{3}{8}e^4,
\end{align}
and $r_{\rm g}$ is the radius of giration. Furthermore $k/T$ depends on the star being convective or radiative. In the case of a convective envelope, $k/T$ is given by \citet[eq. 30-33]{Hurley2002}, based on \citet{Rasio1996}:
\begin{equation}
 \left(\frac{k}{T}\right)_{\mathrm{conv}} = \frac{2}{21} \frac{f_{\mathrm{conv}}}{\tau_{\mathrm{conv}}} \frac{M_{\mathrm{env}}}{M} {\rm yr}^{-1}.
\end{equation}
Where $\tau_{\mathrm{conv}}$ is the eddy turnover time-scale (the timescale on which the largest convective cells turn over):
\begin{equation}
 \tau_{\mathrm{conv}} = 0.4311 \left( \frac{M_{\mathrm{env}} R_{\mathrm{env}} (R - R_{\mathrm{env}}/2)}{3\,L} \right)^{1/3}.
\end{equation}
And $f_{\mathrm{conv}}$ is a numerical factor depending on the tidal pumping time-scale $P_{\mathrm{tid}}$:
\begin{equation}
 f_{\mathrm{conv}} = \min \left[ 1, \left(\frac{P_{\mathrm{tid}}}{2\,\tau_{\mathrm{conv}}}\right)^2 \right],
\end{equation}
\begin{equation}
 \frac{1}{P_{\mathrm{tid}}} = \left| \frac{1}{P_{\mathrm{orb}}} - \frac{1}{P_{\mathrm{spin}}} \right|.
\end{equation}
$M_{\mathrm{env}}$ and $R_{\mathrm{env}}$ are respectively the mass in the outer stellar convective zone, and the stellar radius at the base of that zone. L is the luminosity.

In the case of a radiative envelope, the damping is caused by a range of oscillations that are driven by the tidal field. $k/T$ is then (\citealt[eq. 42-43]{Hurley2002} based on \citealt{Zahn1975, Zahn1977}):
\begin{equation}
 \left(\frac{k}{T}\right)_{\mathrm{rad,d}} = 1.9782 \cdot 10^4 \frac{M_\dsub R_\dsub^2}{a^5} (1 + q_\asub)^{5/6} E_2 \ \ {\rm yr}^{-1}.
\end{equation}
Where $E_2$ is a second order tidal coefficient which can be fitted with:
\begin{equation}
 E_2 = 1.592 \cdot 10^{-9} M_\dsub^{2.84}.
\end{equation}

The total effect on the change in orbital eccentricity is then obtained by combining the tidal forces on both components:
\begin{equation}
 \dot{e}_{\mathrm{tides}} = \dot{e}_{\mathrm{tides,d}} + \dot{e}_{\mathrm{tides,a}}.
\end{equation}

\subsection{Circumbinary disks}\label{app_cb_disks_mesa}
Circumbinary disks (CB disks) can form around binaries during the Roche-lobe overflow phase, if part of the mass can leave the system through the outer Lagrange points and form a Keplerian disk around the binary. The CB disk -- binary resonant and non-resonant interactions have been described by \citet{Goldreich1979} and \citet{Artymowicz1994}, by using a linear-perturbation theory. There are two main assumptions in these models; the first is that the disk is thin (0.01 < $H/R$ < 0.1, where $H$ and $R$ are respectively the thickness and the half-angular-momentum radius of the disk). The second assumption is that the nonaxisymmetric potential perturbations are small around the average binary potential. 

The effect of the disk -- binary resonances on the orbital parameters has been the subject of many studies. In MESA we follow the approach of \citet{Artymowicz1994} and \citet{Lubow1996}. The model of \citet{Lubow1996} for small and moderate eccentricities ($e$ < 0.2) is based on the result of smooth particle hydrodynamic (SPH) simulations, which show that for a disk in which the viscosity is independent of the density, the torque is independent of resonance strength and width. The variation in the orbital separation due to the inner and outer Lindblad resonances is \citep{Lubow1996}:
\begin{equation}
 \frac{\dot{a}}{a} = -\frac{2l}{m} \cdot \frac{J_{\rm D}}{J_{\rm B}} \cdot \frac{1}{\tau_v}. \label{eq_disk_dlna}
\end{equation}
Where $l$ and $m$ are integer numbers indicating which resonance has the strongest contribution. $J_{\rm D}$ and $J_{\rm B}$ are respectively the angular momentum in the disk and in the binary, and $\tau_v$ is the viscous evolution timescale, which is the timescale on which matter diffuses through the disk under the effect of the viscous torque. It is given by:
\begin{equation}
 \frac{1}{\tau_v} = \alpha_{\rm D} \left(\frac{H}{R}\right)^2 \Omega_b. \label{eq_disk_tau_visc}
\end{equation}
Where $\alpha_{\rm D}$ is the viscosity parameter of the disk and $\Omega_b$ is the orbital angular frequency. With a disk viscosity of $\alpha_{\rm D} = 0.1$, the viscous timescale is typically on the order of $10^5$ yr. 

To determine the angular momentum of a Keplerian disk, one need to know the surface mass distribution $\sigma$ in the disk. $J_{\rm D}$ is then given by:
\begin{equation}
 J_{\rm D} = \int_A r \cdot \sigma \cdot v\ {\rm d}A = \int_{\rin}^{\rout} r \cdot \sigma\cdot \sqrt{G M_{\rm B} / r}\ 2\ \pi\ r\ {\rm d}r,
\end{equation}
where the Keplerian velocity of an element in the disk at distance $r$ from the centre of mass, when neglecting the mass of the disk is: $v = \sqrt{G (M_\dsub + M_\asub) / r}$. $\rin$ and $\rout$ are the inner and outer boundaries of the disk.  We assume the surface distribution in the disk to depend on the radius to the power of a free parameter $\delta$:
\begin{equation}
 \sigma(r) = \frac{D_{\mathrm{c}}}{r^{\delta}}. \label{eq_disk_sigma}
\end{equation}
The distribution constant $D_{\mathrm{c}}$ depends on the total disk mass and $\delta$. With this distribution the disk angular momentum can be calculated:
\begin{equation}
 J_{\rm D} = 2 \pi\ D_{\mathrm{c}}\ \sqrt{G (M_\dsub + M_\asub)} \int_{\rin}^{\rout} r^{3/2 - \delta}\ {\rm d}r. \label{eq_disk_Jdisk_int}
\end{equation}
Which depending on parameter $\delta$ has the following solutions:
\begin{equation} 
J_{\rm D} = \label{eq_disk_Jdisk_solved}
  \begin{cases}
    \dfrac{2\pi D_{\mathrm{c}} \sqrt{G (M_\dsub + M_\asub)}}{5/2 - \delta} \left( \rout^{5/2-\delta} - \rin^{5/2-\delta} \right),
      & \delta \neq 5/2.\\
    \\
    2 \pi D_{\mathrm{c}} \sqrt{G (M_\dsub + M_\asub)} \left( \ln{\rout} - \ln{\rin} \right),
      & \delta = 5/2.\\
  \end{cases}
\end{equation}
As shown by the SPH simulations of \citet{Lubow1996}, only the inner part of the disk plays a role in the disk -- binary interactions. Thus in the previous equations the angular momentum is only calculated from the region between $\rin$ and six times the binary separation ($\rout = 6a$). 

To calculate the distribution constant in Eq. (\ref{eq_disk_sigma}), one has to know the total mass in the disk ($M_{\mathrm{disk}}$). The total disk mass can then be related to the distribution constant as:
\begin{equation}
 D_{\mathrm{c}} = \dfrac{M_{\mathrm{disk}}}{2 \pi \int_{\rin}^{\rout} r^{1 - \delta}\ {\rm d}r}.
\end{equation}
Which depending on $\delta$ has two solutions:
\begin{equation}
D_{\mathrm{c}}= 
  \begin{cases}
    \dfrac{1}{2\pi} \dfrac{(2-\delta) M_{\mathrm{disk}}}{\rout^{(2-\delta)} - \rin^{(2-\delta)}},
      & \delta \neq 2.\\
    \\
    \dfrac{1}{2\pi} \dfrac{M_{\mathrm{disk}}}{\ln{\rout} - \ln{\rin}},
      & \delta = 2.\\
  \end{cases}
\end{equation}

The main difference between the implementation here and that of \citet{Dermine2013} is that the disk-surface-distribution parameter $\delta$ in our model is consistent throughout the calculation of the disk angular momentum and the distribution constant $D_{\mathrm{c}}$, while \citet{Dermine2013} used $\delta = 2.5$ in the former, while $\delta = 1$ was used in the latter.

We consider two processes to determine the inner radius of the CB disk. The interaction between the binary and the disk will clear the inner part of the disk. SPH simulations performed by \cite{Artymowicz1994} show that the inner radius depends on the Reynolds number of the disk gas ($\Re = (H/R)^{-2}\alpha^{1}$) and the eccentricity. A fitting formula of their results was derived by \cite{Dermine2013}:
\begin{equation}
 R_{\mathrm{in, SPH}} = 1.7 \cdot \frac{3}{8} \log{(\Re \sqrt{e})} \,\, \mathrm{AU}. \label{eq_disk_rin_sph}
\end{equation}
The second limiting factor on the inner radius is the dust condensation temperature.  Based on \citet[Eq. 1-12]{Dullemond2010}, and adding an offset for the binary separation we obtain:
\begin{equation}
 R_{\mathrm{in, dust}} = \sqrt{ \dfrac{L_\dsub + L_\asub}{4 \pi\ \sigma_{\mathrm{bol}}\ T_{\mathrm{cond}}^4} } + \frac{a (M_\asub L_\dsub + M_\dsub L_\asub)}{(M_\dsub + M_\asub)(L_\dsub + L_\asub)}. \label{eq_disk_rin_dust}
\end{equation}
Where $L_\dsub$ and $L_\asub$ are respectively the luminosity of the donor and accretor, and $\sigma_{\mathrm{bol}}$ is the Stefan-Boltzmann constant. $T_{\mathrm{cond}}$ is the dust condensation temperature which we take at 1500 K. The last factor in Eq. \ref{eq_disk_rin_dust}, is the average binary separation weighted by the luminosity of both components. The actual inner radius of the disk is the maximum of that determined from SPH simulations and the dust condensation radius:
\begin{equation}
 \rin = \max \left[\, R_{\mathrm{in, SPH}}\, ,\, R_{\mathrm{in, dust}}\, \right]. \label{eq_disk_rin}
\end{equation}

There are no direct observations of the outer radius of a CB disk around a post-AGB or post-RGB binary. Post-AGB and post-RGB disks are likely similar to protoplanetary disks, thus it can be assumed that the surface mass distribution of a CB disk does not follow the same behaviour along the whole radius. The inner part of the disk follows $\sigma(r) \sim r^{-\delta}, \delta = 1 - 2$, while the outer part has an exponential decline in $\sigma$. For the CB disk-binary interaction only the inner part of the disk is important. The outer disk radius here then represents the radius where the exponential drop in mass distribution starts, and the maximum disk mass represents the mass in this inner part of the disk. We assume an outer disk radius of $250\,AU$. See for example \citet{Bujarrabal2007, Bujarrabal2013, Bujarrabal2015}. Keep in mind that this outer radius is not the same as the six time the separation radius that is used in calculating the effective disk angular momentum.

The change in eccentricity due to Lindblad resonances can be given in function of $\dot{a}/a$, and depends on the eccentricity. For small eccentricities this can be calculated analytically. In the range $e \le 0.1 \sqrt{\alpha_{\rm D}}$ the $m=l$ resonance dominates, while in the region $0.1 \sqrt{\alpha_{\rm D}} < e \le 0.2$ the $m=2$, $l=1$ resonance is dominant. For small $e$ the analytic form of $\dot{e}$ is \citep[see also][]{Lubow2000}:
\begin{equation}
 \dot{e}_{\mathrm{disk}} = \dfrac{1-e^2}{e + \dfrac{\alpha_{\rm D}}{100 e}} \left( \dfrac{l}{m} - \dfrac{1}{\sqrt{1 - e^2}} \right) \cdot \dfrac{\dot{a}}{a}, 
      \ \ \ e \leq 0.2.  \label{eq_disks_edot_low}
\end{equation}
For average eccentricities the above equation is extrapolated according to an efficiency decreasing with $1/e$:
\begin{equation}
 \dot{e}_{\mathrm{disk}} = \left(\frac{c_4}{e} + c_5\right) \cdot \dfrac{\dot{a}}{a},\ \ \ 0.2 < e \leq 0.7. \label{eq_disks_edot_mid}
\end{equation}
Where the constants are $c_4 = -1.3652$ and $c_5 = - 1.9504$ so that Eq. \ref{eq_disks_edot_mid} smoothly connects to Eq. \ref{eq_disks_edot_low}, while going to $0$ at $e = 0.7$. For larger eccentricities resonances that damp the eccentricity start to dominate, thus $\dot{e}_{\mathrm{disk}} = 0$ in the range $e > 0.7$ \citep[e.g.][]{Roedig2011}.

The change in angular momentum due to disk - binary interactions is:
\begin{equation}
 \dot{J}_{\mathrm{disk}} = J_{\mathrm{orb}} \cdot \left( \frac{\dot{a}}{2{a}} - \frac{e\dot{e}_{\mathrm{disk}}}{1-e^2} \right)
\end{equation}
Where $\dot{a}/a$ is given by Eq. (\ref{eq_disk_dlna}).

The mass feeding the CB disk is the fraction of the mass lost from the binary system through the outer Lagrange point during RLOF. The CB disk mass-loss rate is determined by the maximum mass in the disk $M_{\mathrm{disk,max}}$, and the life time of the disk $\tau_{\mathrm{disk}}$. Both are input parameters in our model. The rate in which the disk loses/gains mass is then:
\begin{equation}
 \dot{M}_{\mathrm{disk}} = \delta\,|\dot{M}_{\mathrm{rlof}}| - \frac{M_{\mathrm{disk,max}}}{\tau_{\mathrm{disk}}}.
\end{equation}
In this implementation the disk will continue existing for a period $\tau_{\mathrm{disk}}$ after RLOF stops.

\end{appendix}

\end{document}